\documentclass[sigconf]{acmart}
\usepackage{multirow}
\usepackage{graphicx}
\usepackage{tabularx} 
\usepackage{booktabs} 
\usepackage{enumitem}
\setlist[itemize]{leftmargin=*}
\newcommand{\sys}{{VizGroup}{}}

\newcommand{\track}{{\textit{Trackers} }{}}
\newcommand{\alert}{{\textit{Alerts} }{}}

\AtBeginDocument{%
  \providecommand\BibTeX{{%
    \normalfont B\kern-0.5em{\scshape i\kern-0.25em b}\kern-0.8em\TeX}}}

\copyrightyear{2024}
\acmYear{2024}
\setcopyright{acmlicensed}\acmConference[UIST '24]{The 37th Annual ACM Symposium on User Interface Software and Technology}{October 13--16, 2024}{Pittsburgh, PA, USA}
\acmBooktitle{The 37th Annual ACM Symposium on User Interface Software and Technology (UIST '24), October 13--16, 2024, Pittsburgh, PA, USA}
\acmDOI{10.1145/3654777.3676347}
\acmISBN{979-8-4007-0628-8/24/10}

\begin{document}
\begin{teaserfigure}
  \includegraphics[width=\textwidth]{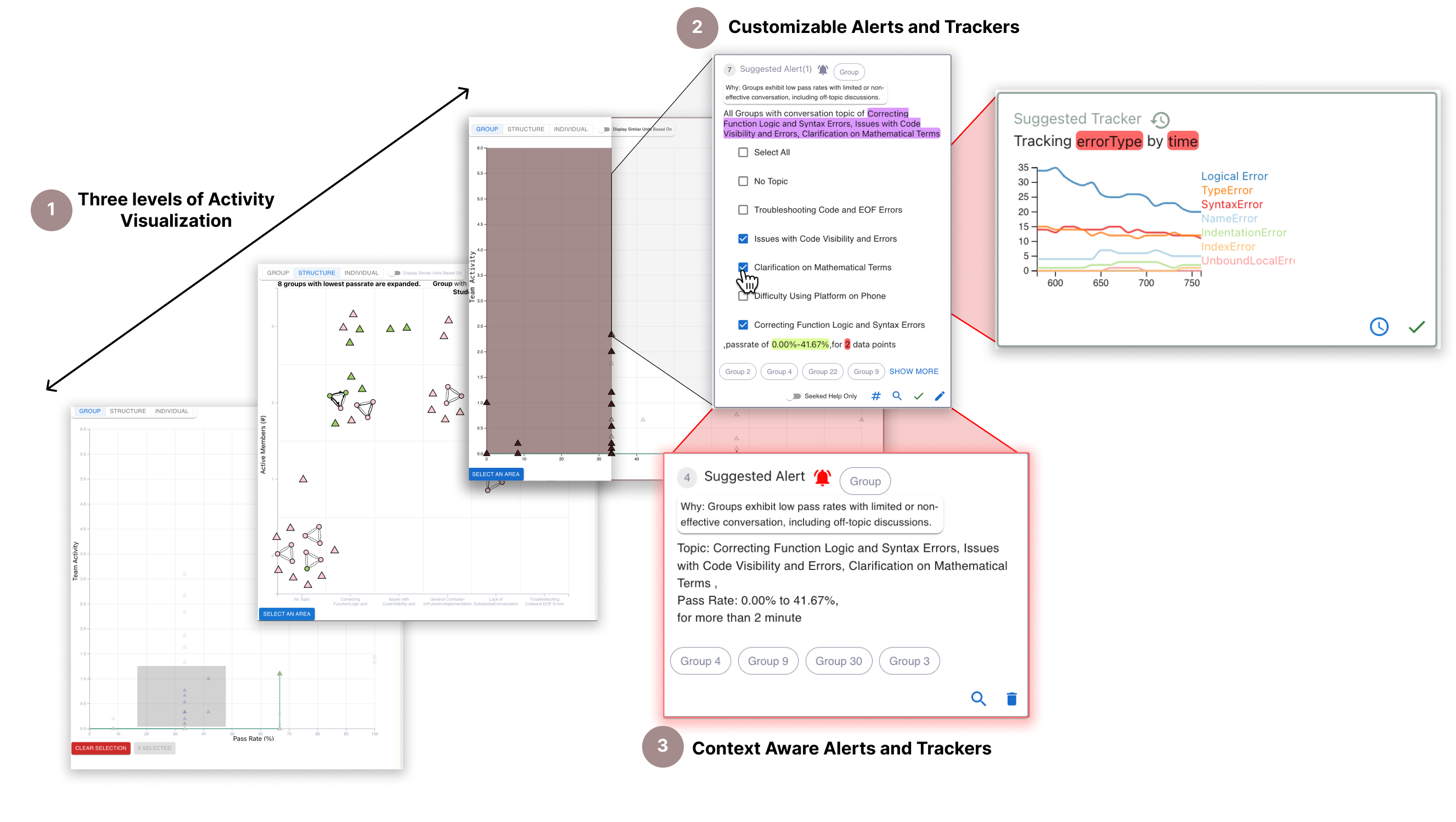}
  \caption{VizGroup, an LLMs-assisted system for real-time collaborative learning analytics, visualizes student performance and collaboration via a dynamic 2D scatter plot and provides proactive notifications for timely interventions.}
  \label{fig:teaser}
\end{teaserfigure}
\title{\sys{}: An AI-Assisted Event-Driven System for Real-Time Collaborative Programming Learning Analytics}

\author{Xiaohang Tang}
\affiliation{%
  \institution{Virginia Tech}
  \city{Blacksburg}
  \state{Virginia}
  \country{USA}
}
\email{xiaohangtang@vt.edu}

\author{Sam Wong}
\affiliation{%
  \institution{University of Washington}
  \city{Seattle}
  \state{Washington}
  \country{USA}
}
\email{samw627@uw.edu}

\author{Kevin Pu}
\affiliation{%
  \institution{University of Toronto}
  \city{Toronto}
  \state{Ontario}
  \country{Canada}
}
\email{kevin.pu@mail.utoronto.ca}

\author{Xi Chen}
\affiliation{%
  \institution{Virginia Tech}
  \city{Blacksburg}
  \state{Virginia}
  \country{USA}
}
\email{xic@vt.edu}

\author{Yalong Yang}
\affiliation{%
  \institution{Georgia Institute of Technology}
  \city{Atlanta}
  \state{Georgia}
  \country{USA}
}
\email{yalong.yang@gatech.edu}

\author{Yan Chen}
\affiliation{%
  \institution{Virginia Tech}
  \city{Blacksburg}
  \state{Virginia}
  \country{USA}
}
\email{ych@vt.edu}

\renewcommand{\shortauthors}{Xiaohang Tang, Sam Wong, Kevin Pu, Xi Chen, Yalong Yang, and Yan Chen}

\begin{abstract}
Programming instructors often conduct collaborative learning activities, like Peer Instruction, to foster a deeper understanding in students and enhance their engagement with learning. These activities, however, may not always yield productive outcomes due to the diversity of student mental models and their ineffective collaboration. In this work, we introduce \sys{}, an AI-assisted system that enables programming instructors to easily oversee students' real-time collaborative learning behaviors during large programming courses. \sys{} leverages Large Language Models (LLMs) to recommend event specifications for instructors so that they can simultaneously track and receive alerts about key correlation patterns between various collaboration metrics and ongoing coding tasks. We evaluated \sys{} with 12 instructors in a comparison study using a dataset collected from a Peer Instruction activity that was conducted in a large programming lecture. 
The results showed that \sys{} helped instructors effectively overview, narrow down, and track nuances throughout students' behaviors.

\end{abstract}
\maketitle

\section{Introduction}

Students' active collaboration with peers while learning can promote engagement, deepen their understanding of concepts, and enhance their problem-solving skills~\cite{smith1992collaborative}. 
In Computer Science Education, peer learning activities such as group discussions~\cite{maloney2008programming}, pair programming~\cite{preston2005pair}, code reviews~\cite{grissom2000practical}, peer instruction~\cite{crouch2001peer}, and peer assessments~\cite{smith2012using} have been employed to foster cooperative learning environments.
Peer Instruction (PI), for example, is an in-class instructional strategy that emphasizes students' active construction of a conceptual understanding with their peers~\cite{crouch2001peer}.  During peer instruction, students first individually respond to a question (i.e., write and submit their code independently), then discuss their code with peers, modify their code, and finally re-submit it. Peer instruction has been shown to be effective at reducing failure rates, improving retention, and enhancing exam performance across various fields, including computer science~\cite{bouvier2019factors, lee2013can, porter2013halving, porter2013retaining, simon2013we}.

Although numerous studies have highlighted the benefits of PI,  
prior work has advocated for guidance during peer collaboration~\cite{umapathy2017meta} and recommended that teaching staff should manage pair interactions in programming labs~\cite{williams2008eleven}.
Yet, it is challenging to design effective management tools to assist instructors in conducting PI activities in large programming classes where large volumes of data about groups can be generated.
We argue that this is largely because the interplay between discussion and learning outcomes during a PI session has received less attention, making it difficult to identify and observe meaningful learner interaction patterns. 
It is unclear if all discussions positively impact overall learning outcomes or if ineffective communication can hinder progress~\cite{porter2011peer}.
By developing tools that enable instructors to better observe and be aware of these interaction patterns, they could begin to understand the relationship between discussion and learning outcomes in the context of their specific courses.

Tools, such as visual analytics (VA), i.e., the combination of automated analysis and interactive visualizations, have shown promise in identifying patterns in large-scale data~\cite{keim2008visual,keim2008visual2,ceneda2016characterizing}. 
 
While most visual analytics (VA) tools support offline analysis of previously collected data, they often lack real-time analysis capabilities~\cite{chen2015peakvizor}. This limits instructors' ability to provide immediate, data-driven interventions that could enhance collaborative performance, particularly when unexpected behavior patterns emerge.
Other systems, such as Groupnamics~\cite{sato2023groupnamics} and Pair-Up~\cite{yang2023pair}, have explored the effects of real-time collaboration analytics, but they did not adequately address the need to analyze the relationship between group collaboration and learning outcomes at scale.
The increasing volume and complexity of data generated during collaborative learning activities can overwhelm instructors, hindering their ability to identify and track events that demand time-sensitive attention. 
Meanwhile, recent advances in Large Language Models (LLMs) have demonstrated the potential to perform real-time data analysis at scale, but it is unclear how to effectively use LLMs to organize and present the information in an intuitive way for collaborative learning analytics.

To investigate the specific design needs and challenges instructors face, we deployed a technology probe in a large programming class (100+ students) where the instructor conducted a PI coding exercise. Our exploration highlighted that (1) instructors need to be able to easily track multiple patterns of correlation between collaboration and the coding exercise in real-time, (2) they need to be informed about emerging patterns in group activity over time, and (3) they need to be able to get a sense of how interaction patterns correlate with the future success of the groups.
These findings underscore the importance of designing systems that support instructors in managing collaborative learning environments by enabling them to monitor multiple types of information without being overly constrained by a prescribed approach.

Based on these findings, we developed \sys{}, an LLMs-assisted system that streamlines the process of overseeing real-time collaborative learning analytics during a programming lecture. \sys{} displays and updates a 2D scatter plot that visualizes collaboration information and students' performance in near real-time. After inputting user interactions (e.g., clicking on a topic) and urgent patterns found in historical data (e.g., groups not chatting after a new code issue occurs) into an LLM, \sys{} will proactively recommend \textit{notifications} (i.e., intelligent monitoring units) that track specific metrics and alert users to important changes or patterns in the data.

To assess \sys{}'s usability and effectiveness, we conducted a between-subject study with 12 participants with teaching experience. Participants used a basic visual analytics tool without a notification system and then used \sys{} with or without our LLM recommendation notifications.
The results showed that compared to a version of \sys{} without the notification recommendation, \sys{} with suggested units helped instructors create additional monitoring units that were previously undiscoverable on their own. 
These recommendations covered a more diverse range of metrics, providing a more comprehensive understanding of the learning process. 
Furthermore, we found evidence that the suggested notifications influenced the participants' decision-making when selecting the following monitoring unit criteria.

Our research makes the following contributions:
\begin{itemize}
    \item Design implications from our formative study that aim to enhance instructors' capacities to monitor and comprehend class-wide collaboration dynamics as they occur.
    \item A new approach that uses contextual information, such as user interactions and real-time data changes, to generate recommendations for tracker and alert creation for novices while using real-time learning visual analytics.
    \item \sys{}, a novel AI-assisted monitoring system for collaborative learning analytics that streamlines the monitoring of key patterns in data via intelligent, context-aware notification creation.
\end{itemize}

\section{Related Work}
 
This section reviews four research fields that inspired our work: collaborative learning, learning analytic systems, collaboration analytic systems, and notification systems.

\subsection{Collaborative Learning}

Collaboration plays a pivotal role in educational policy, research, and technology~\cite{graesser2018advancing}, with teams becoming the model of choice to foster economic competitiveness, improve quality of life, and ensure national security~\cite{fiore2018collaborative}. 
In education, decades of research have shown that collaborative learning can increase student motivation by engaging students in active, hands-on activities (e.g., discussions) that are complemented by immediate peer feedback~\cite{wang2021puzzleme}
Studies have consistently demonstrated that collaborative learning approaches yield greater learning gains than traditional methods, which often rely on passive lectures and standardized exams ~\cite{hake1998interactive, prince2004does}. 
Dowell et al. have developed methods to identify individual roles in collaborative learning contexts to gauge socio-cognitive behaviors~\cite{dowell2019group}.
Additionally, systems like PeerStudio~\cite{kulkarni2015peerstudio} and TalkAbout~\cite{kulkarni2015talkabout} capitalized on peer feedback to improve student performance in MOOCs by strategically connecting peers based on their performance and geographic locations.

Even with such interventions, the efficacy of collaborative learning is not guaranteed. An examination of peer discussions in a large introductory Astronomy course revealed that a significant portion (i.e., 37.7\%) were unproductive, highlighting the common pitfalls of unsupervised conversations~\cite{james2011listening}. Further research into students' help-seeking behaviors has shown that novices often struggle to pose well-formed questions due to their incomplete mental models and lack of the tools to effectively seek help, even when they understand the subject matter~\cite{chen2017codeon, chen2016towards, chen2020edcode}. Unfortunately, instructors may not become aware of these challenges until it is too late, if at all. Therefore, there is a pressing need to track group activities in real-time to enhance the quality of collaborative learning experiences.
\sys{} addresses these critical gaps by providing in-depth analytics of collaborative learning behaviors to enhance the management and facilitation of collaborative dynamics in large-scale, co-located programming learning settings.

\subsection{Learning Analytics}

Learning analytics, an emerging field that focuses on analyzing and visualizing learner data to enhance educational outcomes, provides educators with a fresh perspective on understanding and improving the learning process~\cite{clow2013overview}.
This discipline has been an active research area in the past decade in the HCI community as a result of the substantial growth in the amount of data available about learners, and it is connected to management strategies that emphasize quantitative measures.
Our project builds upon the foundation laid by prior research in learning support and analytics systems, which have been instrumental in enhancing the understanding of students' learning performances. These systems can be broadly categorized into two settings: synchronous and asynchronous.

\textbf{Synchronous Settings}. In real-time analytics, tools like Lumilo~\cite{holstein2018classroom}, VizProg~\cite{zhang2023vizprog}, and Codeopticon~\cite{guo2015codeopticon}, have offered instructors live insights into student activities such as coding and doing math on their computer. 
Other tools such as EduSense~\cite{ahuja2019edusense}, AffectiveSpotlight~\cite{murali2021affectivespotlight}, and Glancee~\cite{ma2022glancee} assessed student and audience facial or body gestures to provide new analytics to instructors. 
These studies found that providing instructors with rich data about their students' statuses can reveal key insights about their level of engagement, which helps instructors better manage the challenges of large class sizes. 
These tools, however, only focus on students' individual activities and did not consider the collaborative aspects of live classroom settings. 

\textbf{Asynchronous Setting}. In contrast, asynchronous learning analytics systems like Overcode~\cite{glassman2015overcode}, Foobaz~\cite{glassman2015foobaz}, and MistakeBrowser and FixPropagator~\cite{head2017writing} have excelled at generating personalized feedback for clusters of student submissions that displayed similar patterns. 
Although beneficial for learning, these systems primarily concentrated on individual student achievements without considering the broader social context of learning.

\subsection{Collaboration Analytics}

The HCI and Education communities have advocated for collaboration analytics systems that are theoretically grounded, adaptively capture comprehensive interaction data, model collaboration sensitively to context, respect users ethically, and provide customized support catering to the distinct characteristics of individuals and groups~\cite{schneider2021collaboration}. 
Instructors, however, often face challenges while monitoring and guiding group discussions as they may not have access to adequate information about a group and its members or are unable to constantly facilitate conversations.

Prior work, such as Pair-Up~\cite{yang2023pair}, explored collaboration analytics by examining the transition behaviors of K-12 students from individual to group learning. Groupdynamics~\cite{sato2023groupnamics} provided a summary of the vocal activities and statuses of up to 10 small discussion groups. While these systems provided valuable insights into collaboration dynamics, they focused on prescribed metrics and lacked support for discovering the emerging correlation patterns between collaboration and the tasks that people are working on. 

Prior work~\cite{chamillard2011using} and our formative study (Section \ref{FS}) have suggested the need to support the correlation between working tasks and collaboration metrics. 
Moreover, existing collaboration analytics tools often lack guidance on how instructors should prioritize their attention. 
Researchers have suggested a more flexible notification system that enables users to customize the group behaviors they wish to monitor and receive alerts about, such as group status. Additionally, keeping track of multiple visited groups and their descriptions is mentally taxing for instructors~\cite{sato2023groupnamics}.
\sys{} aims to fill these gaps by providing instructors with the ability to analyze the correlation between collaboration dynamics and the tasks students are working on, and to easily create notifications and alerts about insightful patterns.

\subsection{Notification Systems}
Notification is an important feature in today's software systems. It allows users to receive relevant updates more timely. Common examples include monitors and alert functions in medical systems that are designed to support doctors in receiving time-critical decision support~\cite{mastrianni2022alerts, mastrianni2022pop}, and email and message notifications in mobile applications that are designed to help users stay aware of personal matters~\cite{li2023alert}. There were also works on providing real-time notifications and feedback in classrooms to enhance teachers' awareness~\cite{martinez2014mtfeedback} and support teachers' orchestration~\cite{tissenbaum2019supporting}. However, these works often focus on the impact that notifications have on users' work and well-being, such as work disruption, and the appropriate time to receive notifications~\cite{auda2018understanding, ho2018nurture}. In contrast, we look at how to help users choose and create notifications to optimize their resources or effort in real-time.

Recent LLMs may help process this massive number of multi-faceted data in real-time. However, it remains unexplored how to effectively leverage LLMs in notification systems that support users in identifying urgent information to monitor for allocating help-seeking resources. Our work takes inspiration from prior notification systems and fills this gap by exploring the use of LLMs to generate context-aware notifications in collaborative learning environments, enabling instructors to efficiently manage and support students' learning experiences.

\begin{figure*}[h]
    \centering
    \includegraphics[width=1\linewidth]{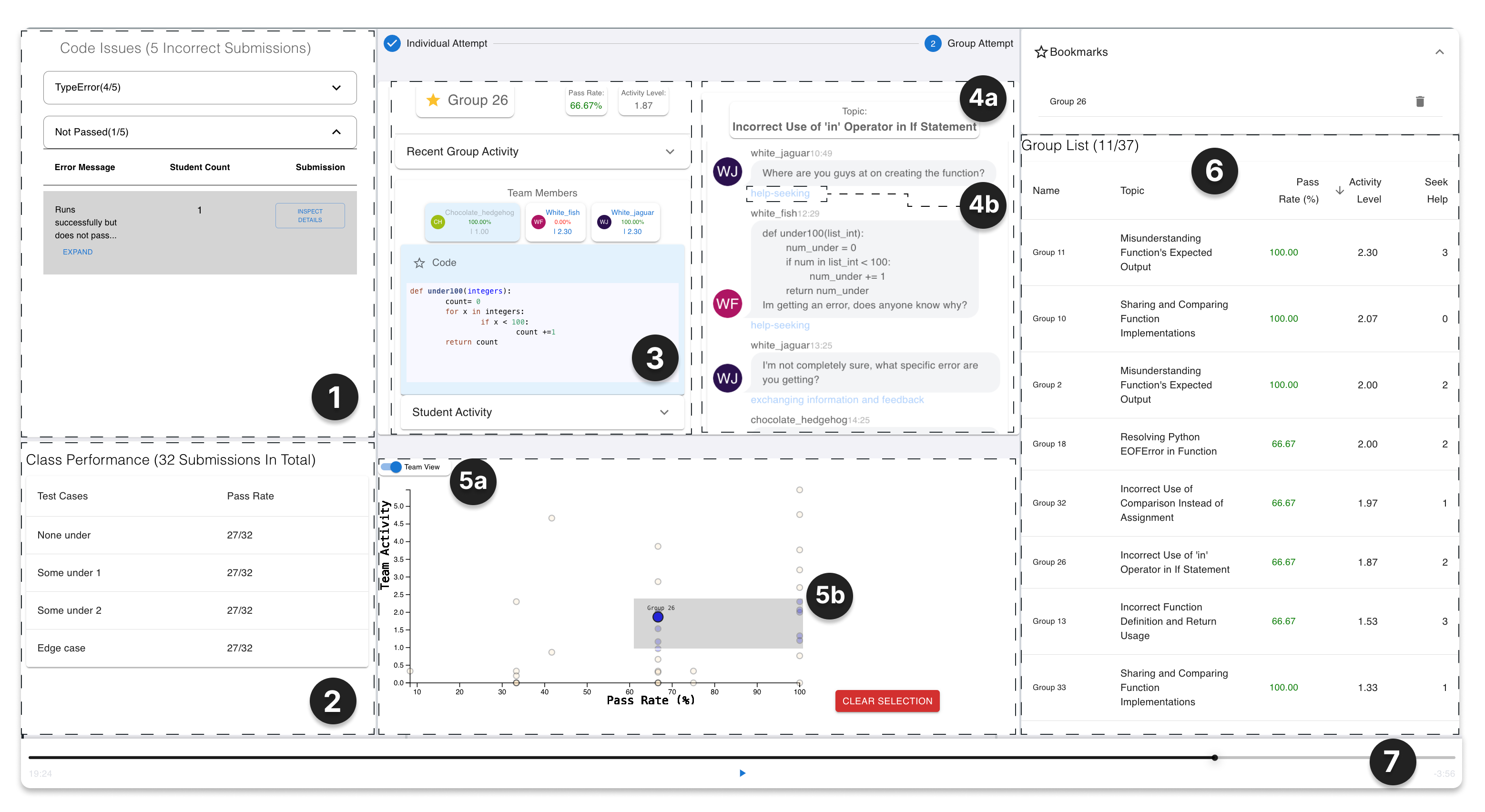}
    \caption{The technology probe's user interface. (1) Code issue list aggregated code errors based on student code submissions. (2) the class performance of the number of student passing each unit test. (3) A student activity panel that showed the activity level and pass rate of each group and group member. The chat panels contained the chat history of the selected group, (4a) a summary of the conversation and (4b) the tagged chat message (in blue) as summarised by the LLM. (5a) The scatter plot that visualized students' activity level and could be toggled between individual view and student view. (5b) Instructors could examine multiple data points by highlighting a region and (6) each data point would be displayed in the Group List panel. (7) Each session could be reviewed via playback controls.}
    \label{fig:probe}
\end{figure*}

\section{Formative Study}
\label{FS}
Due to the lack of real-time large-scale collaborative learning VA systems, we deployed a technology probe \cite{hutchinson2003technology} to explore the potential informational needs of instructors and the corresponding design challenges. 
To discover design considerations and derive interface designs for our probe, we first conducted semi-structured interviews with four experienced instructors of large programming courses at our institution.
These need-finding interviews focused on the types of information instructions desired in real-time during the group discussion activity and the decisions they aimed to make based on this information. 
We found that instructors desired to discern patterns in code and group discussions and their interplay in how discussions aided students in resolving their issues.

\subsection{Probe System Design}
In response to interview findings, we developed a probe VA system that collected student code submissions and evaluated their correctness using unit tests. It used an LLM to summarize students' chat messages in group activities and presented the information using an individual and a group-level visualization, along with students' task progress and errors.

To analyze the student discussions in each group, messages were tagged individually and the conversation was summarized.
To define meaningful patterns, we draw upon principles from social interdependence theory ~\cite{johnson2013cooperation}, which posits that effective collaborative learning is characterized by \textit{promotive interactions} that occur as individuals encourage and facilitate each other's efforts to reach the group's goals (such as maximizing each other's learning).
Based on the theory, all chat messages were tagged by GPT-4 using 6 categories:
1. Help-giving, 2. Help-seeking, 3. Exchanging information and feedback, 4. Joint reflection on progress and process, 5. Mutual encouragement and challenging, 6. Not related to the class. 
GPT-4 tagged each message in real-time and displayed the tag below each chat message (Fig.\ref{fig:probe}.4b). Furthermore, the main topic discussed in each group conversation was summarized by GPT-4 every time a new message was sent in the chat (Fig.\ref{fig:probe}.4a), thus enabling instructors to obtain an overview of the conversations among different groups without having to read all the chat messages.

To enable instructors to efficiently track information in real-time, the probe included a live visualization of individual and group levels (Fig.\ref{fig:probe}.5a).
On an individual level, the probe calculated students' activity scores based on their chat history. Each message from categories 1-5 contributed 1.0 to the activity score, and each category 6 message contributed 0.3 to the score.
For the group level, the score was computed as the average activity score of all group members.
The activity score of each group was represented as a scatter plot dot, where the y-axis was the activity level of the group and the x-axis denoted the unit test pass rate.
Participants could manually select a dot to view detailed student or group information or select a range to track multiple students or groups (Fig.\ref{fig:probe}.5b).

\subsection{Methodology}
We recruited 8 participants with experience teaching programming courses to participate in our within-subject study where they evaluated the probe and a baseline system.
Participants then completed a survey and interview.
We captured an in-class peer instruction session from a large-scale introductory university programming course (i.e., 111 students, 37 groups) and used its live playback during the study.
The baseline condition was an ablated version of the probe without the group discussion visualization and intelligent features like discussion topic summary, team activity level, and message tags (i.e., without 4a, 4b, 5a, and 5b in Figure~\ref{fig:probe}), similar to existing educational VA systems.

For each condition, participants were asked to complete four tasks (Appendix \ref{formative_study_task}), designed to evaluate instructors' understanding of student and group progress and dynamics based on the insights from the need-finding interviews.
Each task required participants to experience a segment of the recorded instruction session.
Three tasks involved identifying patterns in a group discussion and students' help-seeking using the visualization system. One open-ended task asked participants to identify any issue they found important as instructors. The system conditions and tasks were counterbalanced.

\begin{figure*}[ht]
\centering
\includegraphics[width=\textwidth]{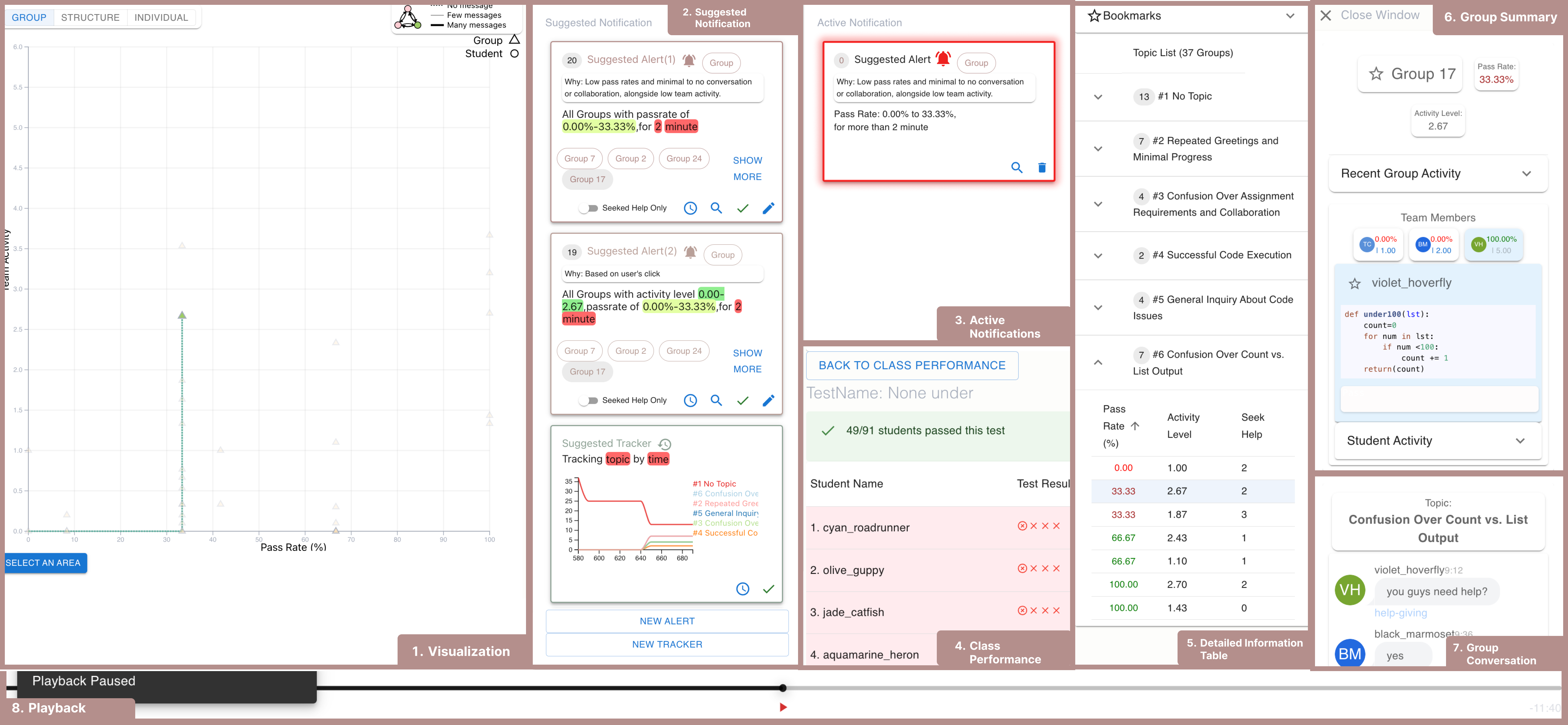}
\caption{\sys{}'s User Interface. (1) Scatterplot displays student's progress via \textit{group view}, \textit{group structure view} and \textit{individual view}. (2) Suggested notification panel that displays context-aware notification suggestions and user-defined notifications. (3) Active notification panel that displays active trackers and alerts. (4) Class performance that shows the number of students that passes each unit tests.
(5) List of Details on group conversation topic and student error messages. (6) Group information panel that shows group and individual activity. (7) LLM summary of student conversation and conversation history.  (8) Each session can be controlled with playback.}
~\label{fig:\sys{}}
\end{figure*}

\subsection{Results and Design Considerations}
To analyze the results, we calculated the correctness of the identified trend (i.e., a particular student's or group's task pass rate and discussion increased or decreased over time), and the precision and recall of the classification tasks (i.e., \textit{identify all the students that have engaged in group discussions over a period of time}).
For Task 1, a T-test found no significant difference in the precision ($p = 0.11$) and recall ($p = 0.44$) between the two conditions.
For Task 2, participants had a higher recall using the probe (Probe: $Mean = 0.990, \sigma = 0.029 $; Baseline: $Mean = 0.742, \sigma = 0.200, p < 0.01$), while the precision was on par for both conditions ($p = 0.96$).
We also found that participants had better accuracy while tracking and understanding group dynamics trends during Task 3 while using the probe ($Mean = 0.875, \sigma = 0.173$) than the baseline system ($Mean = 0.667, \sigma = 0.178, p < 0.05$).
We did not find a significant difference in the time spent completing the tasks (Probe: $Mean = 1277.29 s$, Baseline: $Mean = 1244.86 s$, $p > 0.05$).

We also coded the issues participants recorded during the open-ended quiz task (Task 4), and distinguished between two types of descriptions: general and detailed. 
Participants identified more issues overall with the probe ($Mean = 3.000, \sigma = 0.930$) than with the baseline ($Mean = 1.875, \sigma = 1.130, p < 0.05$). 
The results also showed that participants identified more detailed issues using the probe ($Mean = 2.000, \sigma=1.200$) than the baseline ($Mean = 0.500, \sigma = 0.760, p < 0.01$). 
Specifically, with the baseline system, 4 participants described issues in a general way such as \textit{"Syntax Error"}, and 1 participant could not describe any issue. 
While with the probe, 7 participants were able to explore both the code and the discussions, recording detailed issues such as \textit{"returning not an int or wrong int from calculation"}.

From the post-study survey and interviews, we found that the probe's ability to present \textit{discussion topics} and \textit{activity levels} prompted participants to investigate patterns starting from group engagement down to student-level coding challenges. 6 participants (P1, P3, P5, P6, P7, P8) reported discussion topic as one of the \textit{"most satisfying"} features because it is \textit{"very useful for instructor to find the common issues"} without \textit{``checking it one by one''} (P6). P4 also suggested that activity levels addressed the issue of \textit{``a little bit nervous (seeing) automatically generated topics''.}
Moreover, participants reported that the probe's \textit{high-level information display} and \textit{visualization} features \textit{``provided about what was going on in the class.''} (P5) and \textit{``allow me to quickly overview the status of the students, and help me easily identify the student who is struggling''} (P2). These features streamlined the process of monitoring and analyzing collaborative learning dynamics, thereby facilitating a more nuanced understanding of student and group performance over time.

However, the probe demanded higher \textit{cognitive loads} as participants needed to manually keep track of an identified pattern over time, making the observation unscalable to monitor multiple patterns (e.g., students who were not engaging, students who lacked support, etc.). P6 stated that \textit{``sometimes I forget some of the students' performance before that I can't tell if the pass rate was increased.''}
Participants also mentioned the importance of focusing on a \textit{``level or window''} when inspecting students and groups \textit{``that can make it more efficient as we are not distracted by other information''} (P5).
In addition, different patterns were relevant in \textit{different contexts}. For example, initially, students may not be engaged in a discussion as they work on a coding task, but this idleness could become an issue once everyone else has submitted their code.
The context dependency prompts a design that would enable contextually-aware pattern discovery and monitoring in the next iteration of our system design.

\section{Design Goals}

Our formative study revealed two primary effects driving our DGs: (1) instructors used different levels of information scope to understand student progress, and (2) instructors were overloaded when tracking progress and providing support. Based on these findings and the key challenges identified in prior work, three design goals guided the iterative development of \sys{} to support instructors in being able to easily monitor collaborative learning in large programming classes.

\begin{itemize}
    \item \textbf{DG1. Efficient and contextualized navigation across different levels of granularity.}
    To enable instructors to identify patterns of concern in real-time using interactive learning analytics, the navigation process must be both efficient and adaptable to different levels of granularity, allowing seamless transitions between individual, group, and class-wide perspectives.
    \item \textbf{DG2. Real-time and attention-free monitoring.}
    To address the overload instructors experienced when trying to follow the progress of parallel groups, there is a need for a system that offers real-time, attention-free monitoring, along with mechanisms to selectively guide their attention and minimize interference with their workflow.

    \item \textbf{DG3. Guiding instructors' attention while considering both instructors' and students' needs in context.}
    To avoid compounding instructors' overload with unnecessary interruptions, mechanisms for guiding instructors' attention (e.g., via notifications) must be context-aware, taking into account the needs of both instructors and students.
\end{itemize}
With the three design goals in mind, we iteratively revised our probe system to enhance instructors' ability to navigate, track, and set alerts to understand real-time collaborative programming learning patterns.
The resulting \sys{} user interface has three main panels (Figure~\ref{fig:\sys{}}): (1) a three-level view of collaborative learning behaviors that offers insights into collaboration at varying levels of abstraction; (2) a notification panel that display user- and AI-suggested notifications; and (3) an information table that lists metrics related to team activities and coding performance. 
These panels are interconnected through data-binding, thus ensuring that selecting a subset of data in one panel will automatically update the corresponding views in the other panels.

\begin{figure*}
\centering
\includegraphics[width=\linewidth]{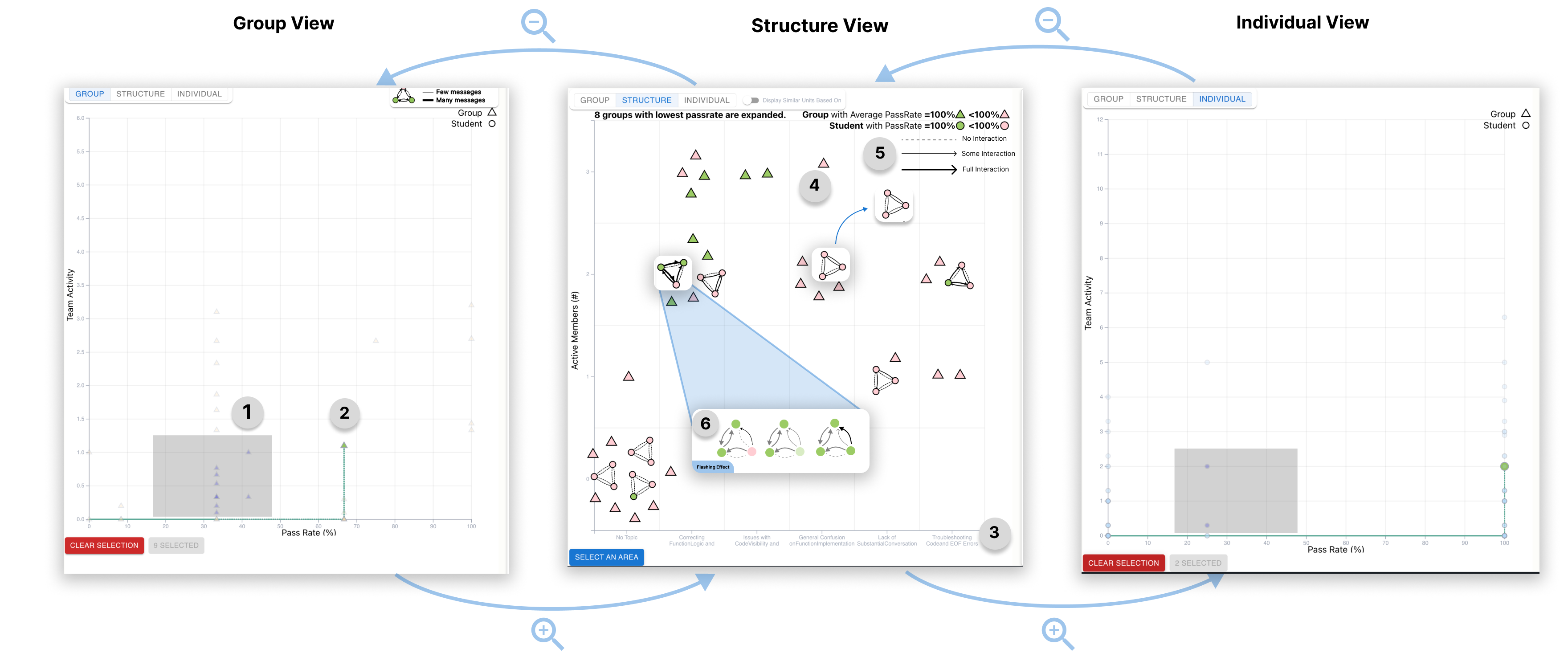}
\caption{Overview of collaborative learning visualization with Group, Structure, and Individual Views, switchable by zooming. Features: (1) Area highlight for inspecting multiple points, (2) History trace of data points upon selection, (3) Structure View plots participants against conversation topics, (4) Group data point location changes with topic or participation shifts, (5) Arrows indicate participation levels, (6) Flashing effect for changes in pass rates or interactions.}
~\label{fig:Three Level View}
\end{figure*}

\section{\sys{}}
\sys{} was implemented as a web-based visualization tool using React and D3.js for its core functionality and OpenAI's GPT-4\footnote{\url{https://platform.openai.com/docs/models/gpt-4-and-gpt-4-turbo}
} and text-embedding-3-large\footnote{\url{https://platform.openai.com/docs/models/embeddings}} for specific features. 

As \sys{} expanded on the probe VA system, it shared similar features including (1) a chat panel that utilized the LLM to analyze group and individual activity (Figure ~\ref{fig:\sys{}}.7 and ~\ref{fig:probe}.3), an LLM summary of group discussion topics (Figure ~\ref{fig:\sys{}}.6 and ~\ref{fig:probe}.4a,b) and a class performance panel (Figure ~\ref{fig:\sys{}}.4 and ~\ref{fig:probe}.2). We also modified the detailed information table to display an aggregated list of conversation topics when the Group view was selected on the scatterplot or an aggregated list of student code errors when the Individual view was selected on the scatterplot. 

During a class session, \sys{} recorded data on an individual student and on a group level (Table ~\ref{tab:my-table}). On an individual level, \sys{} tracked each student’s pass rate, activity level, and type of code errors. On the group activity level, it tracked the average pass rate of the group, the activity level of the group, the group's conversation topic, and the team structure of the group. We chose these five dimensions because of their critical importance, as highlighted by previous research~\cite{yang2023pair, sato2023groupnamics, 10.1145/3657604.3662025/CFlow}, and based on the preferences and experience of instructors from the formative study and the data we have. \textit{Pass Rate} serves as a proxy for students' progress in understanding and utilizing the learning objectives, while \textit{Activity Level} indicates students' engagement in collaboration. Likewise, \textit{Team Structure } reflects students' participation at a group level. \textit{Code Issue} highlights errors as identified by the compiler in their submissions, and \textit{Conversation Topic} reveals what students encounter in their collaborations. 

\subsection{Three Level View of Collaborative Learning}
To facilitate the visualization of different levels of detail during collaborative learning (DG1), \sys{} uses a three-level view encompassing high-level group performance and activty (i.e., Group View), the mid-level group interaction structure (i.e., Structure View), and low-level individual performance and activity (i.e., Individual View). Based on the findings from the preliminary studies, each view was designed to reveal specific patterns and insights about the collaborative learning environment and highlight key behavioral analytics that were identified in our research. 

\begin{figure}[b]
\centering
\includegraphics[width=\linewidth]{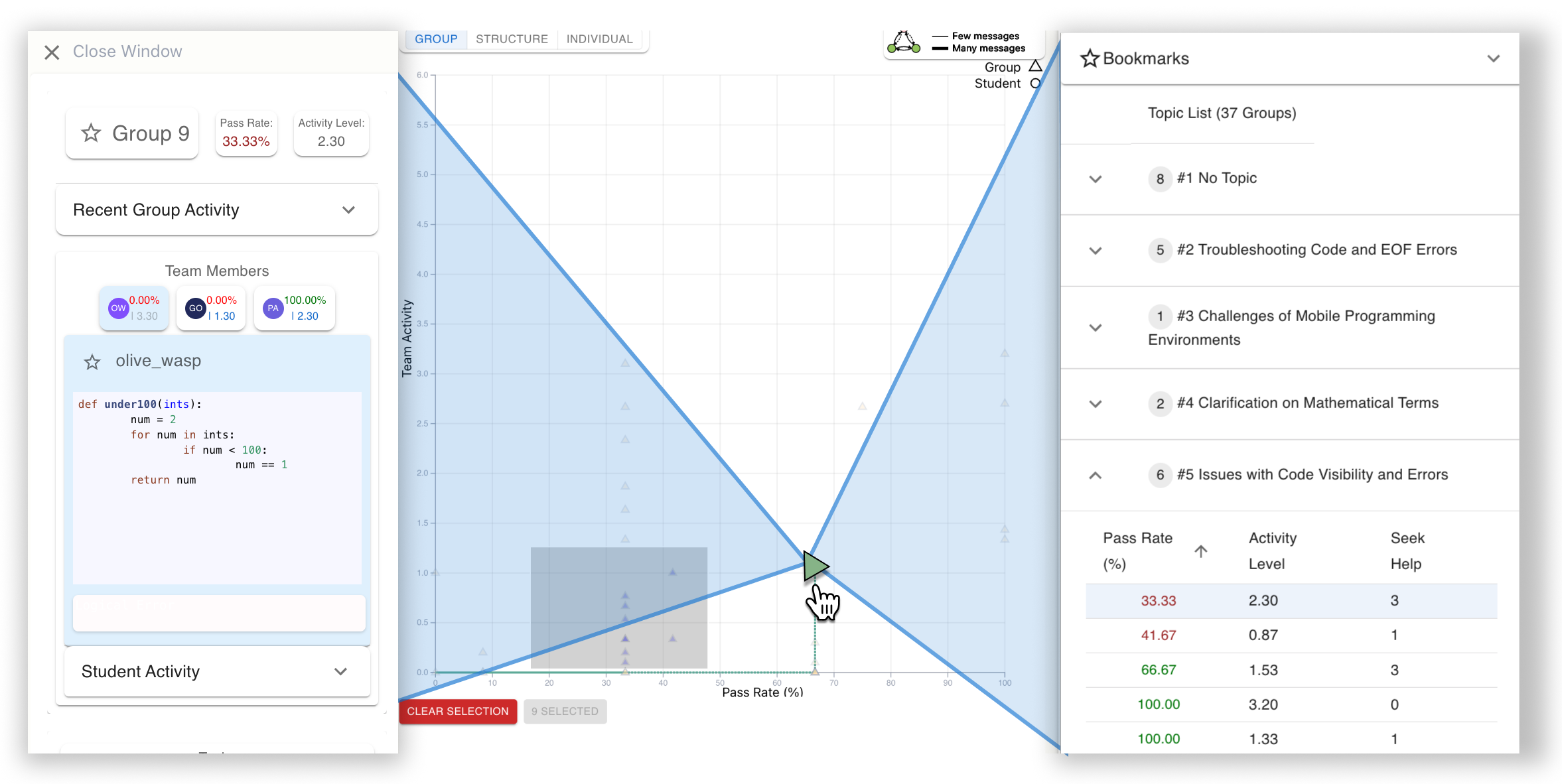}
\caption{Clicking on a group data point displays its information summary panel, and activity and performance table.}
~\label{fig:Click}
\end{figure}

\subsubsection{Group View:} This view presents an aggregated perspective of the group's collaborative interactions, enabling instructors to gauge the average group activity level (y-axis) and average group pass rate (x-axis) over time (Figure ~\ref{fig:Three Level View}). By aggregating data over collective activities, this view facilitates an understanding of group dynamics and collaborative patterns at a glance. 


\begin{table*}[t]
\begin{tabularx}{\textwidth}{lllXX} 
\toprule
                   & Student & Group & Description                                                            & Range \\
\midrule
Pass Rate          & \checkmark       & \checkmark     & The percentage of correct unit test passed by each student or group      & [0\%, 100\%] \\
Activity Level     & \checkmark       & \checkmark     & The level of participation in a group discussion based on chat frequency & [0.0, 12.0] \\
Code Issues        & \checkmark       &       & The code error message in each student submission                          & \textit{No Compiling Error, Type Error, Name Error, Indentation Error, Index Error, Syntax Error} \\
Conversation Topic &         & \checkmark     & Each group's topic summary based on their existing chat messages. Individual conversation topics is tagged by LLM (Figure ~\ref{fig:probe}.4b) but is not used in visualization and notifications.            & Varies, example topics include \textit{Correcting Function Logic and Syntax Errors}, \textit{Troubleshooting Code and EOF Errors} \\
Team Structure     &         & \checkmark     & The number of students active in each group                                & [0,3] \\
\bottomrule

\end{tabularx}
\caption{The student activity attributes that are tracked in \sys{}.}
~\label{tab:my-table}
\end{table*}

\subsubsection{Structure View:} This view maps the current topics of discussion (x-axis) against the number of active members in a group (y-axis), offering a detailed snapshot of group dynamics (Figure ~\ref{fig:Three Level View}.3). 
We implemented a topic modeling approach to assess the relevance of each chat message and to distill the overarching subjects of discussion. Initially, we utilized an LLM to generate text embeddings for each conversation, which serve as a nuanced representation of the discussions' content. These embeddings were then clustered using the K-means++~\cite{Bachem_Lucic_Hassani_Krause_2016} algorithm, a method chosen for its efficiency and reliability in grouping semantic word representations~\cite{10.1145/3613904.3642117}. For each identified cluster, we produced a concise summary that captures the essence of the conversations within. To ensure a streamlined and coherent compilation of conversation topics, we compared new summaries with existing ones. If a newly generated summary closely matched any of the previously established summaries, we classified it under that pre-existing category.

We integrated graph representations and incorporated indicators of activity levels directly within the view to simplify the analysis process for instructors. As the activity levels or the discussion topics change, each group cluster data point will transition to a different position of the visualization (Figure ~\ref{fig:Three Level View}.4). Each group's communication pattern is also depicted using arrows to indicate the sender and the recipient of chat messages, with the arrow's thickness representing the level of activity amongst group members (e.g. no messages \includegraphics[height=1em]{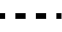}, few messages \includegraphics[height=1em]{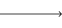},many messages \includegraphics[height=1em]{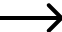 }; Figure ~\ref{fig:Three Level View}.5). The thickness of the arrow is calculated by: 
\[
\min(A \times 0.25, 2) + 1
\]
where A is the activity level of the individual.
The green dots indicating that students passed all unit tests, whereas the pink indicated they did not fully pass all unit tests. When there were updates to the group conversation topics or the number of active members, a flashing animation was applied to the changed data point to catch the instructor's attention (Figure ~\ref{fig:Three Level View}.6).
To prevent information overload, \sys{} displayed up to eight group structures with the lowest pass rate simultaneously. Expanding the lowest performing groups would enable instructors to identify if group activity contributed to the low pass rates. 
\begin{figure}[b]
\centering
\includegraphics[width=\linewidth]{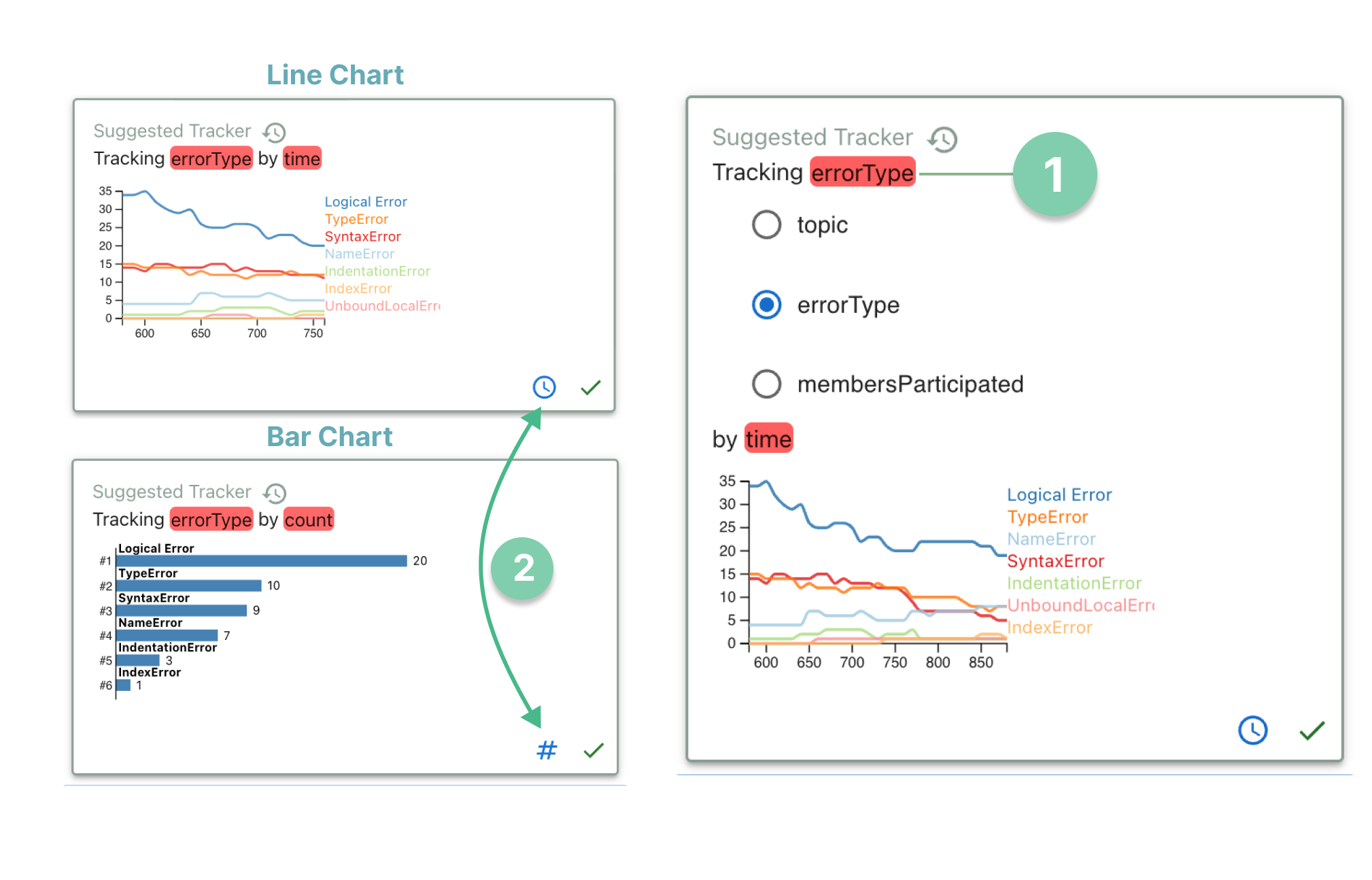}
\caption{Trackers visualizes student activity data using bar charts and line charts. (1) Instructors choose the student activity attribute to be tracked and (2) can switch between the two visualization.}
~\label{fig:notify-layout}
\end{figure}

\subsubsection{Individual View:} This view enables instructors to assess each student's engagement (y-axis) and performance (x-axis) closely (Figure ~\ref{fig:Three Level View}). It was designed to highlight individual behaviors, such as students who might be struggling or feeling excluded, in relation to their peers' progress. This view aids instructors in quickly identifying students in need of additional support or those who are excelling and can potentially serve as resources for their peers.

\begin{figure*}
\centering
\includegraphics[width=\textwidth]{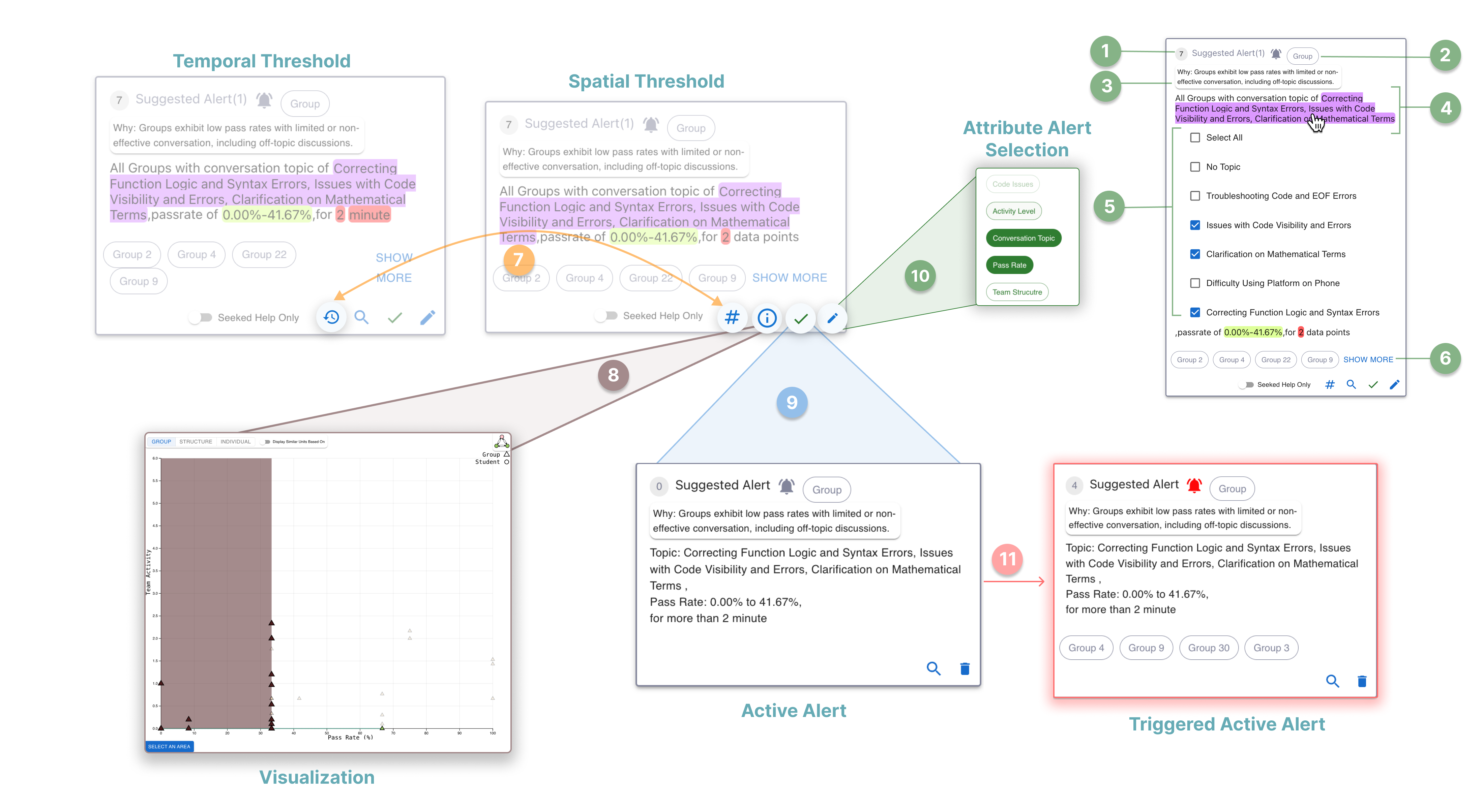}
\caption{Alerts notify instructors if students/groups meets the defined criteria. (1) Count of data points/groups that matches specified criteria, (2) Alert type (group or individual), (3) Notification creation reason, (4) Trigger criteria (modifiable via highlighted area click), (5) Criteria selection drop-down menu, (6) Clickable current data points meeting criteria, displayed on scatter plot and detail panels.(7) Instructors can change between using a spatial threshold or a temporal threshold. (8) Instructors can preview data points on the scatter plot. (9) An alert activation confirmation is sent to the list of Active Notifications. (10) Instructors can select the attributes to be included in the threshold. (11) Flashing Animation when alert is triggered.}
~\label{fig:alert}
\end{figure*}

\subsubsection{Interaction}
To support seamless navigation across these views (DG1), \sys{} enables instructors to zoom in and out. 
Selecting a region (Figure ~\ref{fig:Three Level View}.1) or clicking on a data point provides further insights in the detailed information list and the group details panel (Figure ~\ref{fig:Click}). Clicking on a data point will also show the trace history of the selected data point, which enables instructors to view changes in performance over time (Figure ~\ref{fig:Three Level View}.2).


\subsection{Notifications for Observing Student Activity}
Inspired by Fluid UI, where an influx of information is structured automatically to reduce the gulf of evaluation~\cite{poupyrev2023UltimateInterface}, we designed a structured notification system that enables instructors to create and customize notifications that alert them about key changes in student activity that might require attention and intervention. Specifically, we introduce the concepts of \textit{Tracker} and \textit{Alert} to help instructors to observe changes in student progress. 

Notifications are displayed as two columns: Suggested Notifications (Figure ~\ref{fig:\sys{}}.2) and Active Notifications (Figure ~\ref{fig:\sys{}}.3).Instructors can add and edit existing trackers or alerts in the Suggested Notification panel before activating it. After activation, the notification will appear in the Active Notifications panel, where it will start tracking real time data and push alerts based on instructor-specified criteria. 

\begin{figure}
\centering
\includegraphics[width=1\linewidth]{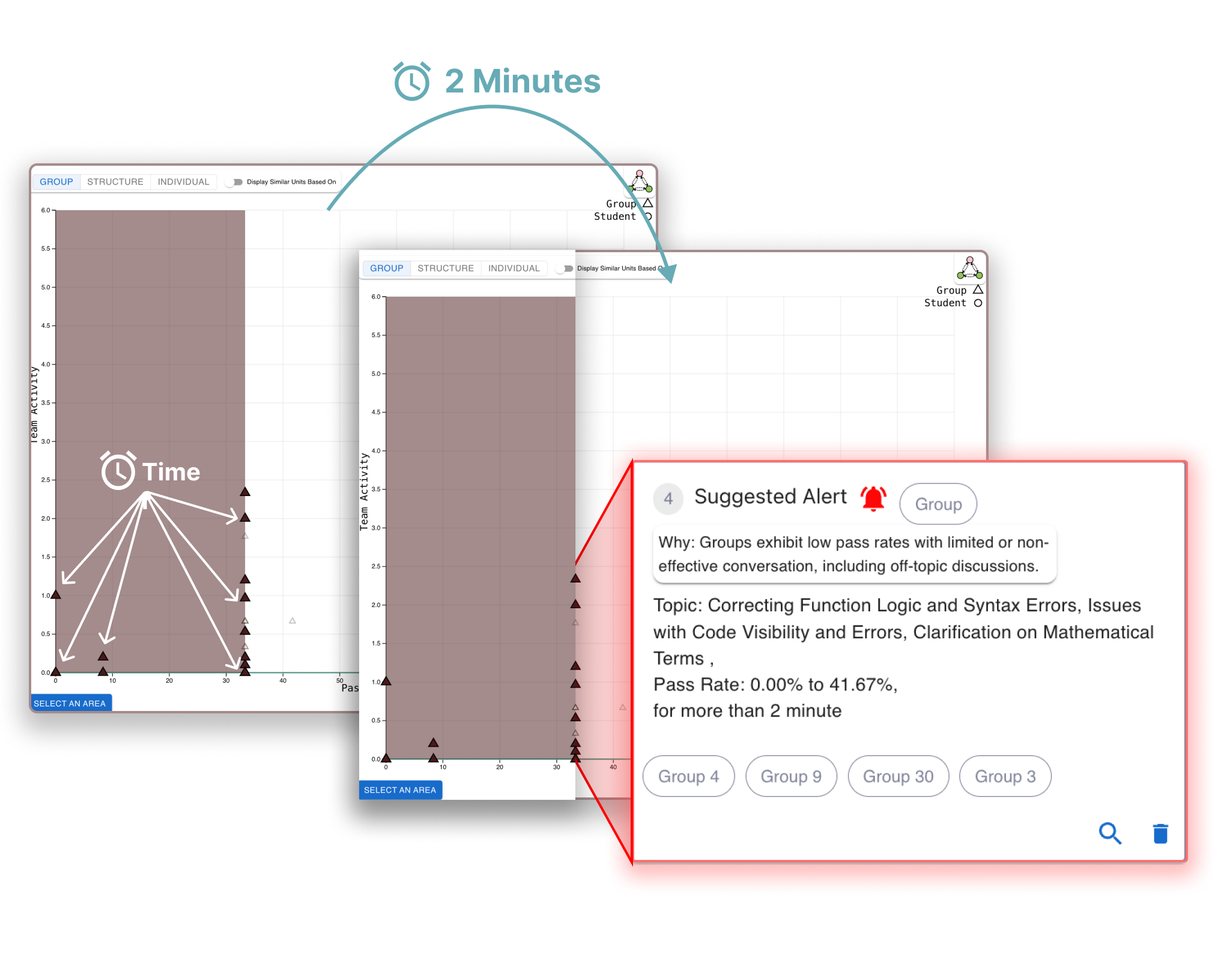}
\caption{Temporal Alert mechanism in 2-Dimension. Data points which remained in the highlighted area for more than 2 minutes will be reported.}
~\label{fig:temporal alert}
\end{figure}

\subsubsection{Trackers (Figure ~\ref{fig:notify-layout})}

\track are visualizations that display the current count of student activity attributes. Instructors can select a variable to be tracked by clicking on the highlighted attribute, which will reveal three options: Code Issues, Conversation Topics, and Members Participated (Figure ~\ref{fig:notify-layout}.1). They can display the visualization as a bar chart that shows the count of each values within a specific attribute (e.g., Code Issues are grouped by the count of different errors, Conversation Topics are grouped by the count of different groups of conversation summaries whereas Members Participated is grouped the the total number of active participants in the group). They can also view a time-series line plot that displays the changes in the different groups of the selected attribute from the start of the session until the current time (Figure ~\ref{fig:notify-layout}). Instructors can toggle between the two visualization easily (Figure ~\ref{fig:notify-layout}.2).

\begin{figure}
\centering
\includegraphics[width=\linewidth]{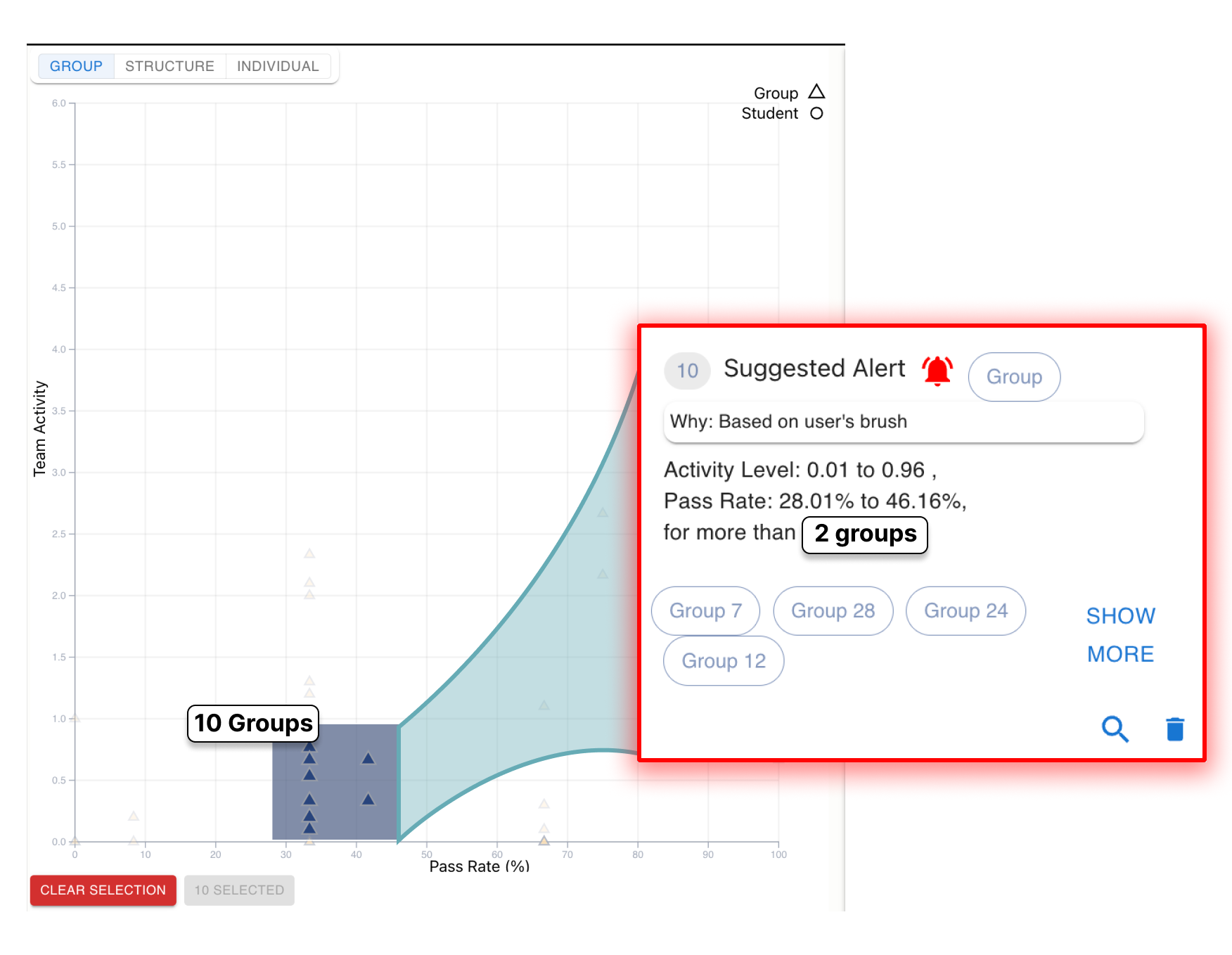}
\caption{Spatial Alert mechanism. \sys{} will notify instructors when the number of data points in selected area (10 data points) exceeds defined threshold  (2 data points).}
~\label{fig:spatial alert}
\end{figure}

\subsubsection{Alerts (Figure ~\ref{fig:alert})}
Drawing from previous research on the use of visual aids to signal the current status of groups in online breakout rooms and alerts that monitors abnormal activity in complex systems~\cite{sato2023groupnamics, bunch2004multiagent}, \alert function as a tool for tracking when student behaviors or group interactions surpass limits set by instructors.
They encapsulate the activity level, group conversation topic, the collective pass rate, team structure for group analysis, and code issues, activity levels, and individual pass rates for individual analysis (Table ~\ref{tab:my-table}). However, 
the alert card shows the alert type (group or individual) (Figure~\ref{fig:alert}.2), the reason why the alert is created (Figure ~\ref{fig:alert}.3), the current criteria for the alert (Figure ~\ref{fig:alert}.4), the list of data points that currently satisfies the criteria (Figure ~\ref{fig:alert}.6) and the total number data points in the list (Figure ~\ref{fig:alert}.1). Instructors can modify the types of activity attributes they want to be alerted about (Figure ~\ref{fig:alert}.10). For each attribute, clicking on the highlighted area reveals a drop down menu (Figures ~\ref{fig:alert}.4), allowing instructors to modify its threshold values (Figure ~\ref{fig:alert}.5). 

Alerts can be activated spatially or temporally. For spatially, the alert is activated when the quantity of students surpassing a predetermined threshold exceeds \textit{n} (e.g., \textit{alert instructors when the number of students that has a pass rate lower than 50\% and an activity level from 0 to 3 is more than 10}; Figure ~\ref{fig:spatial alert}). 

These alerts are based on a 5-dimensional model of student activity attributes. Instructors receive notifications when the number of students that falls under the predefined student activity range, allowing for timely interventions.

\begin{figure}
\centering
\includegraphics[width=1\linewidth]{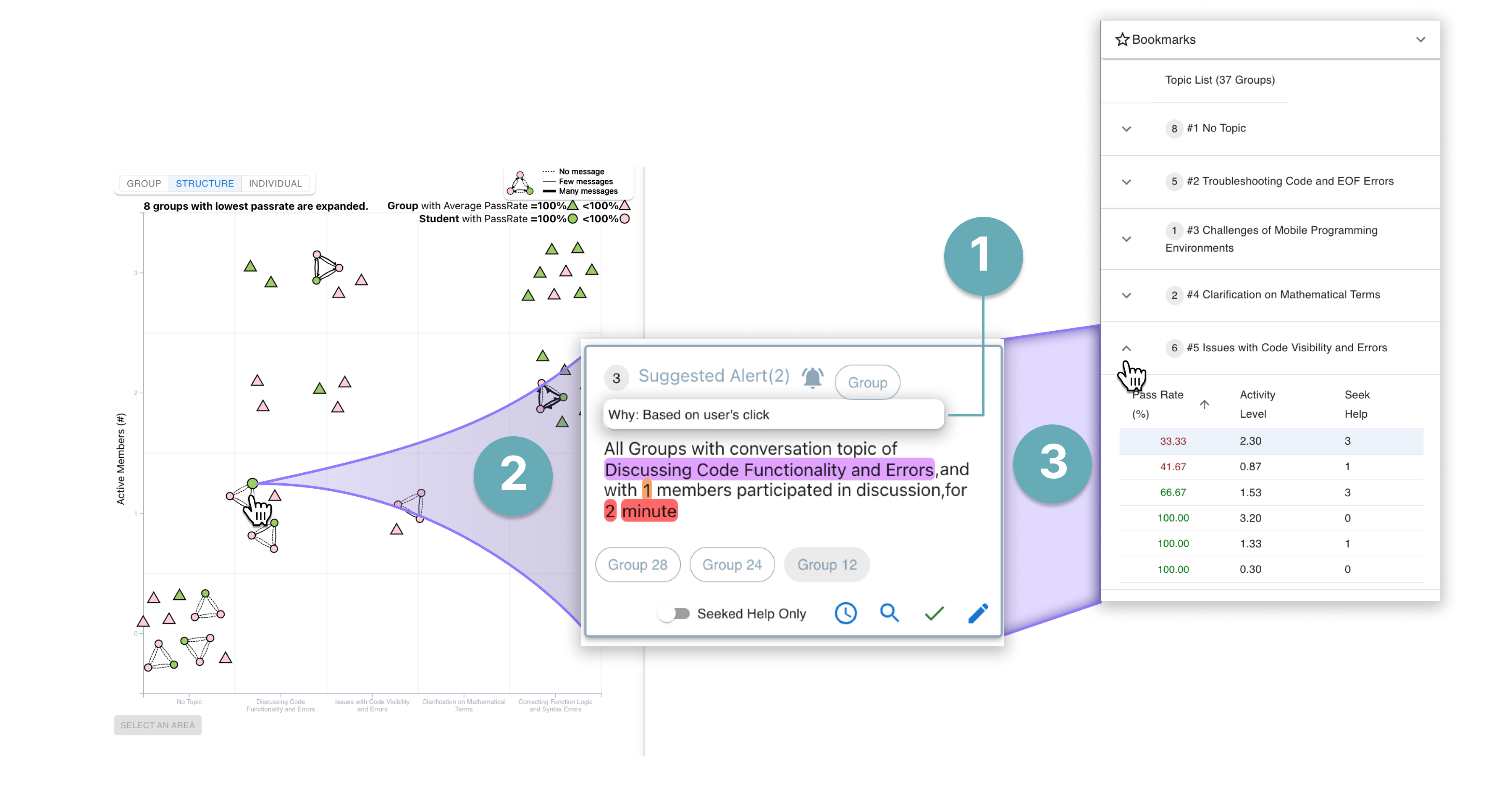}
\caption{User interaction with the UI creates suggested alerts, displaying user clicks as the main reason for notification generation (1). Interacting with data points in Structure View suggest alerts based on the conversation topic and the number of active members in the group (2). Interacting with the detailed list of group topics would suggest topics based on the expanded drop down (3).  }
~\label{fig:ui-suggest}
\end{figure}

Temporally alerts are activated when specific groups or students surpass the predetermined threshold for a duration exceeding \textit{t} seconds or minutes (e.g., \textit{alert instructors of all students that discuss correcting function logic and syntax errors in their group conversation and have a pass rate between 0\% and 41\% for over 2 minutes}; Figure ~\ref{fig:temporal alert}). Consider setting an alarm clock, where a clock will sound when the time is up. Alerts use a similar idea where each datapoint that meets the user defined criteria has an internal alarm clock, keeping track the amount of time they meet the defined criteria. When the time is up for each individual datapoint, namely the period of time they meet the criteria exeeds the user defined time, it will push an alert which will display the datapoint to the instructor for further investigation. 
After defining the alert criteria, a curated list of groups and students that meets the threshold enables instructors to examine associated group details (Figure ~\ref{fig:alert}.6). Moreover, \sys{} provides an option for instructors to preview and visualize data points that currently align with alert criteria, offering a macroscopic view of the tracked groups' spatial distribution (Figure ~\ref{fig:alert}.8). By clicking on the time or number icons (Figure ~\ref{fig:alert}.7), instructors can transition between these alert types.

After confirming the alert thresholds by clicking the green check mark, the alert transitions to an Active Notification (Figure ~\ref{fig:alert}.9). The alert then invokes a flashing animation to ensure that instructor is aware of any updates (Figure ~\ref{fig:alert}.11).
\begin{figure*}
\centering
\includegraphics[width=\textwidth]{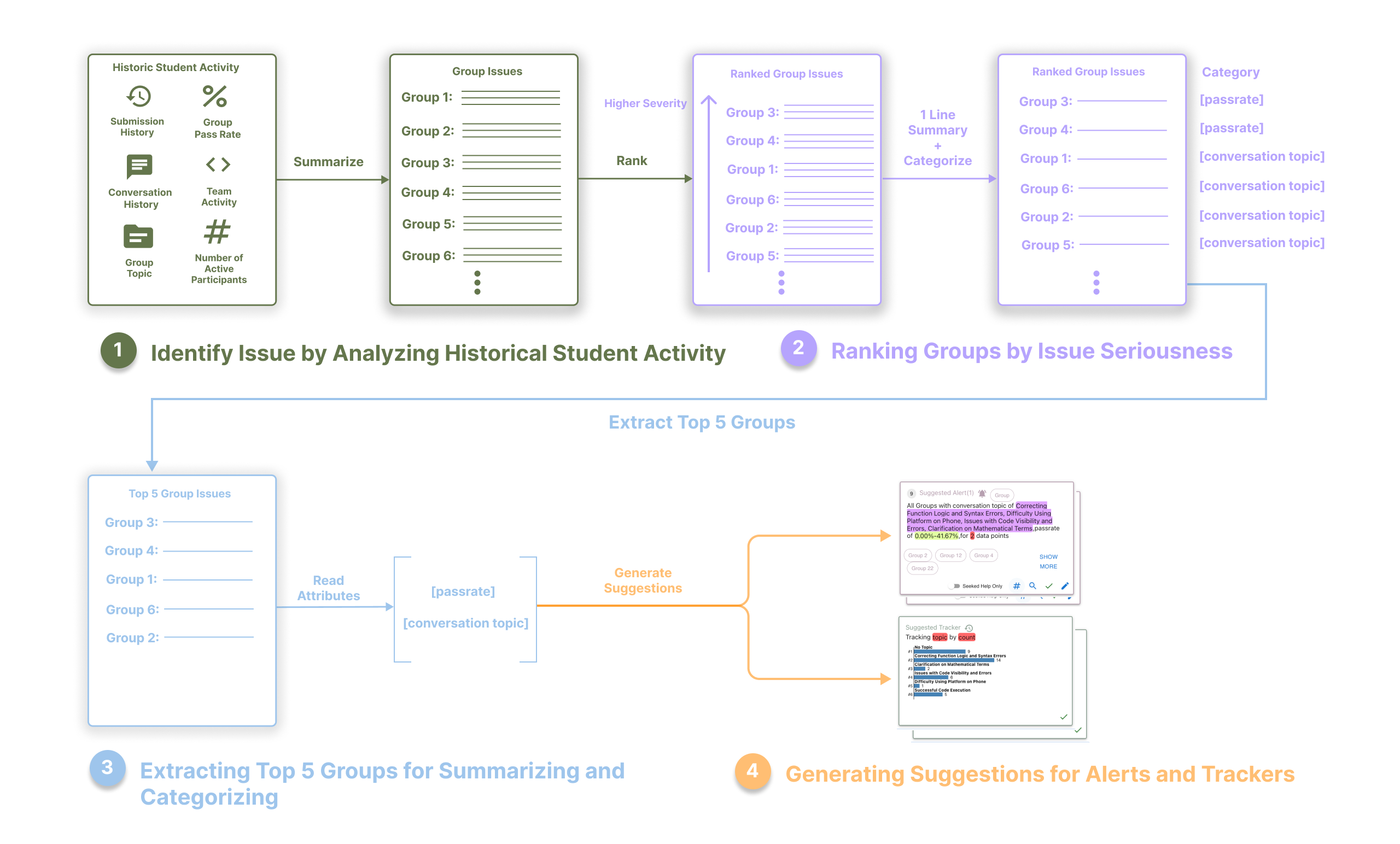}
\caption{Workflow for generating Suggested Notifications.}
~\label{fig:alertprocess}
\end{figure*}

\subsection{Context-Aware Suggestions}
With the rapidly changing nature of student behavior during a large-scale activity, it could be challenging to monitor and create notifications. Thus, \sys{} generates suggested alerts and trackers as templates to help instructors identify key patterns in student activity that might be hard to notice by observing the scatter plot and topic list of details. Suggested Notifications were generated based on user interaction with the \sys{} interface and historic changes in student activity data

\subsubsection{User Interaction}
\sys{} generates suggested notifications for the scatter plot view currently in use. Specifically, group alerts will be suggested when instructors are in group and structure view and individual alerts will be suggested when instructor is in individual view. 
Instructors could also create notifications by selecting areas of interest on the scatter plot. When they click on a specific data point, \sys{} derives a suggestion on student activity attributes to be tracked based on the displayed view of the scatter plot. 
For the Structure View, it would suggest alerts based on the selected group conversation topics and the number of team members active in the group (Figure ~\ref{fig:ui-suggest}.2). 
For the Group and Individual view, it would suggest alerts based on group or individual activity level and pass rate. 
When the instructor highlights an area in the scatter plot, the x and y range of the selected area will be automatically applied towards the suggested notification, enabling instructors to select data points to keep track of through direct manipulation. 
Instructors can use such suggestions as templates to further modify alerts to suit their needs. Similarly, when instructors inspect the rows of the aggregated topics and errors in the detailed information table, suggestions are based on the expanded row (Figure ~\ref{fig:ui-suggest}.3). 
\subsubsection{Historic Student Activity Data}
When instructors are tracking more than 3 attributes that can't be easily represented on a 2D visualization or creating notifications that take into account historic student activity, they may neglect to consider recent student activity that might be relevant. Hence, \sys{} leverages an LLM that takes historic student activity data into account while generating alert suggestions for instructors. Suggestions for individual student and group data are processed separately, and depending on the current view, group or individual related alerts will be suggested. Here we outline the steps the LLM used to generate suggested alerts using historical student activity data. The steps for suggesting group and individual notifications are similar (see Figure~\ref{fig:alertprocess}), so we outline the process of suggesting group notifications for the sake of simplicity (see all used prompts in Appendix A).

\paragraph{Step 1: Identify Issues By Analyzing Historical Student Activity Data}
Following prior work on using LLM to identify challenges in students' learning and collaborations~\cite{10.1145/3636555.3636905predictingchallenge}, we asked the LLM to analyze each group's history of submission attempts, group conversation logs and current status data such as group pass rate, team activity, the summarized group topic, and the number of members participating in the group discussion. After this holistic evaluation of group performance, we asked the LLM to identify specific issues each group was facing. Some examples of issues were no active conversation despite submission attempts or group members asking for help but the conversation lacked problem solving. 

\paragraph{Step 2: Rank Groups by Issue Seriousness}
After identifying issues across groups, we asked the LLM to rank each group based on the severity of the issue, with major issues ranked higher and minor issues ranked lower. Here a high severity issue is defined as an issue that might require instructor and teaching assistant intervention, while a low severity issue can be handled by students in the group. Then, the LLM summarized the identified issue into one sentence for each group. Based on prior work that used LLMs to perform recommendation tasks, we adopted a list-wise ranking approach to rank the identified issues to achieve a balance between cost and performance ~\cite{Dai2023ChatGPTRS}. The LLM also noted the most problematic issue out of their pass rate, the relevance of their conversation to the task, their topic of conversation, and their participation levels during discussions. 

\paragraph{Step 3: Use the Top 5 Groups for Summarization and Issue Categorization}
We then used the top 5 groups with the most severe issue and asked the LLM to summarize the common problems found across these 5 groups. Based on this summary, the LLM then identified which aspects of pass rate, the relevance of their conversation to the task, their topic of conversation, and their participation levels during discussions were present in the summary. 

\paragraph{Step 4: Generating Suggestions for Alerts and Trackers}
From the extracted groups/students, we calculated the range of categorical and numeric data for the alerts. For categorical data, we obtained the set of all values found in the extracted groups/students. For numerical data, we obtained the global minimum and maximum across the 5 groups/students. For all annotated topics summarized in Step 2, we then aggregated the count based on the pass rates, the relevance of their conversation to the task, the topic of conversation, and participation levels during discussions in the summary, and tracked the most frequent categorical data. 

The primary objective of utilizing LLMs in the suggestion generation process is to identify complex patterns in students' code and text-based discussions, transforming these into easily inspectable suggestions to support instructors' learning analytics. To enhance accuracy, we also adopted state-of-the-art prompting engineering techniques such as few-shot prompts~\cite{NEURIPS2020_1457c0d6} and AI-chains~\cite{10.1145/3491102.3517582}.

\section{System Evaluation}
We conducted an in-person user study to examine \sys{}'s usability and effectiveness. We also investigated participants' experiences when using \sys{}'s suggestion feature by including a condition of \sys{}'s ablated version in the study.
\subsection{Participants}
We recruited 12 participants (5 females and 7 males) who had experience teaching programming courses at four-year universities via personal networks, local mailing lists, and snowball sampling. During the study, participants were asked to interact with \sys{} under different conditions to inspect student behavior that was collected in a large programming course at our institute.
Each participant was compensated with \$25 USD for their time and effort.
\subsection{Protocol}
\subsubsection{Live Simulation} During the study, participants watched live playback of the session to simulate a real-time, in-class peer instruction session. To ensure the data participants interacted with was authentic, we used 
real data captured from a large-scale introductory level university programming course's peer instruction session that contained 111 students that were divided into 37 groups. 
The grouping strategy was to gather students who passed the test with those who did not and ensure each group had at least one student who passed the test. During the peer instruction process, students accessed the system through their laptops. They were not able to see group members' real names and their code submissions. The programming exercise for the session was an introductory Python problem to count the number of elements under 100 in a given list of numbers. 

\subsubsection{Conditions}
We used a mixed study design that incorporated both within-subject and between-subject methods, where each participant used the system under two of the following three conditions:
\begin{itemize}
    \item \textbf{Baseline (A):} a baseline version of \sys{} without any intelligent or automatic features such as notifications, topic summaries, and team activities (i.e., without \textit{(2)}, \textit{(3)}, and topics in \textit{(5)} and \textit{(7)} in Figure~\ref{fig:\sys{}}. The system still contained the interactive visualization.
    \item \textbf{\sys{} without suggested notification (B2): } an ablated version of \sys{} without the context-aware suggestion feature. This version still had the notification feature, but all notifications needed to be manually created.
    \item \textbf{\sys{} (B1):} a full AI-assisted version of \sys{} with all its features. Notification suggestions are dynamically generated in \textit{B1}. Interaction-based suggestions were triggered by users' interaction, while the system displayed a new historic-based suggestion every 15 seconds. 
\end{itemize} 
\begin{figure*}
    \centering
    \includegraphics[width=\textwidth]{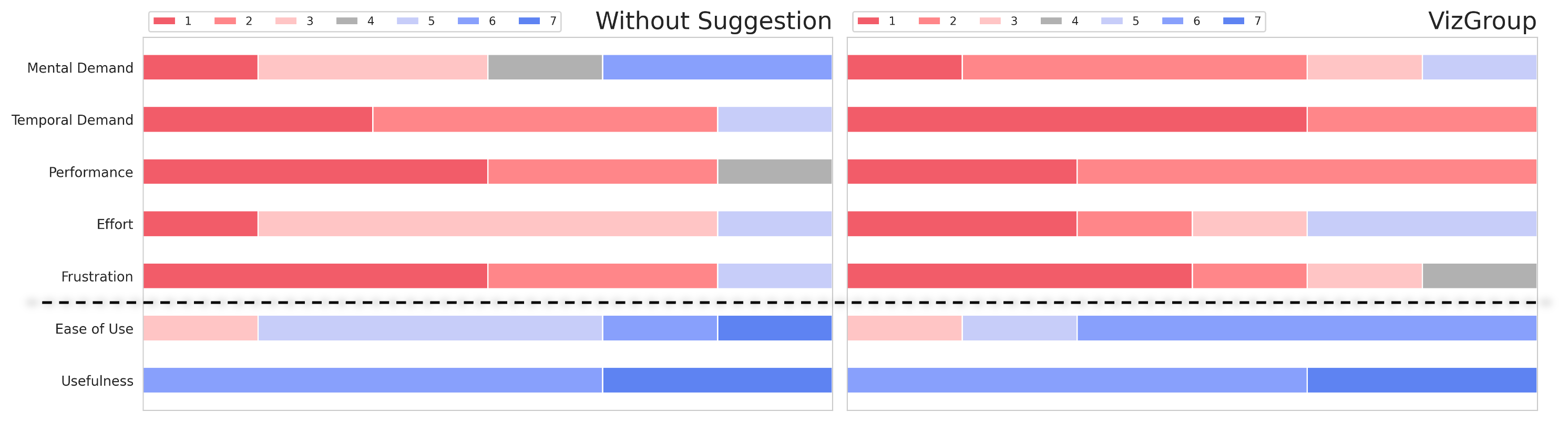}
    \caption{Survey responses after the open-ended tasks. For the NASA-Task questions ~\cite{hart1988development} (up), 1 indicated very low levels of mental demand, temporal demand, successful performance, effort, and frustration. For the ease of use and usefulness questions (down), 1 was very negative and 7 was very positive.}
    \label{fig:likert_bar}
\end{figure*}
\subsubsection{Tasks}
All participants completed two types of tasks:
\begin{itemize}
    \item Quiz. Each quiz contained two classification questions:
    \begin{itemize}
        \item Q1: Identify students who have not passed the test and only sent irrelevant messages in the chat.
        \item Q2: Identify groups with specific difficulties and track those who were stuck for two minutes.
    \end{itemize}
    \item Open-ended task. Each participant was asked to create notifications to identify students or groups that needed a TA to help them while monitoring the class for two minutes.
\end{itemize}
The quiz questions were designed to evaluate the system's ability to enable participants to overview, narrow down, and track nuances throughout the class. The open-ended task sought to evaluate participants' experiences with the customized usage of the notification feature to track students' behavior and groups' interaction over time.
These questions were derived from our formative study and iterated on.
While participants watched the playbacks from one session under both conditions, we used the same clips from the playback for open-ended tasks (between-subject) and used different clips with counterbalanced order for the quiz questions (within-subject) to reduce overhead costs (e.g., getting familiar with an exercise).
We also counterbalanced the conditions to reduce potential learning effect.

\subsubsection{Study Procedure}
At the beginning of each study session, the study coordinator collected informed consent from the participant. Then, the study coordinator gave an explanation of the context of the data and the tasks used in the study. Following this, participants completed two quizzes under two assigned conditions after watching the tutorial of the corresponding system and warming up. After that, participants were asked to work on the open-ended task with the condition equipped with the notification feature. Additionally, they completed a short survey with Likert scale questions and participated in a semi-structured interview after each quiz and open-ended task. Participants were asked to think aloud during the open-ended task and the surveys. Each session took around 60-75 minutes. All study sessions were screen- and audio-recorded.

\subsection{Results}

\subsubsection{Quantitative Results}
We combined the results from the two conditions (\textit{B1, B2}) with the notification features as the quiz questions did not involve the use of the historic-based suggestions (Table \ref{tbl:quiz_results}).
The results showed that participants who used \sys{} had shorter times, higher recall, and higher precision on Q1, and significantly shorter time and higher recall on Q2.

In the open-ended tasks, participants using \sys{} received 18.26 ($\sigma = 1.21$) notification suggestions on average. Among all the suggestions, there were 13 types of notifications categorized by activity attributes while 11 of them were accepted by participants.
Participants created more notifications for both students (\sys{}: $Mean = 2.67, \sigma = 1.37$, \sys{} without suggestion: $Mean =2.00, \sigma =1.26$) and groups (\sys{}: $Mean = 3.00, \sigma = 2.10$, \sys{} without suggestion: $Mean =2.50, \sigma = 1.64$). In total, there were $25.93\%$ more notifications created using \sys{} (\sys{}: $Mean = 5.67, \sigma = 3.44$, \sys{} without suggestion: $Mean =4.50, \sigma =2.81$). There was not a statistically significant difference in the number of created notifications. Of 5.67 created notifications using \sys{}, 2.50 ($\sigma = 2.95$) of them are from historic-based suggestions, 2.67 ($\sigma = 1.21$) of them are from interaction-based suggestions, and 0.50 ($\sigma = 1.22$) are manually created. 
\begin{table}[t!]
    \centering
    \small
    \begin{tabular}{l  c c c  c c c c }
    \toprule
     \multirow{2}{*}{Condition}& \multicolumn{3}{c}{Q1}& & \multicolumn{3}{c}{Q2}\\
     \cline{2-4} \cline{6-8}
      & Time &Prec.  & Recall &   & Time &Prec.  & Recall\\
      
    \midrule
    Baseline & 348 & 0.72 & 0.73 & &  338&0.97  & 0.82\\
    \sys{} & \textbf{224}$^{*}$& \textbf{0.98}$^{**}$ & \textbf{0.96}$^{**}$ & & \textbf{251}$^{**}$&\textbf{1.00} & \textbf{0.98}$^{**}$\\
     \bottomrule
    \end{tabular}
    \caption{Quiz Performance (Time in seconds). The best results for each condition is in bold. $*$ indicates $p < 0.05$, while $**$ indicates $p < 0.01$. }
    \label{tbl:quiz_results}
\end{table}
\subsubsection{Context-Aware Suggestions Facilitated Awareness of Unexpected Patterns}
Based on our observation and the think-aloud process, we found that participants (P3, P9, P11) not only adapted suggested notifications so that they aligned with their mindsets, but also accepted those different from their initial strategy. For instance, P3 stated \textit{``I care about the pass rate, not the type of code issues''} at the beginning of the notification creation phase. However, this participant's mind was changed after inspecting the suggested alert that contained criteria about a \textit{Code Issue}. The participant accepted the suggested criteria since the selected students within the notification \textit{``makes sense''}.
Moreover, in the following monitoring process, the participant created more notifications for \textit{Code Issue}, which was the criterion the participant did not want to pay attention to at the start.

We also noticed that participants exhibited a low level of mental demand ($Median = 2.00, \sigma = 1.38$) using \sys{}, while those participants who used \sys{} without suggestion gained a medium level of mental demand ($Median = 3.50, \sigma = 1.94$) to complete the open-ended tasks (Figure \ref{fig:likert_bar}).

\subsubsection{Context-Aware Suggestions Increased the Diversity of Notifications}
During the open-ended tasks, the notifications created with \sys{} were $35.14\%$ more diverse than the ablated version of \sys{} (Number of notifications containing different types of criteria--\sys{}: $Mean = 4.50, \sigma = 2.66$, \sys{} without suggestion: $Mean=3.33, \sigma=1.63$; not statistically significant). 
For instance, before inspecting the suggested notification, P3 only created a notification about \textit{Pass Rate} and \textit{Activity Level}, however, after inspecting and adapting suggested notifications, P3's notifications covered all of the criteria.

In contrast, P5 created 3 notifications for students using \sys{} without suggestion. However, among these notifications, the only difference was in the categories selected for \textit{Code Issues}. P5 followed the same scope during the notification creation process, which was how most participants (P2, P4, P5, P8) commonly worked with \sys{} without suggestion.

\subsubsection{\sys{} Aided Instructors In Identifying and Tracking Patterns and Work in Parallel}
For Q2, most participants who used \sys{} checked other groups' behaviors, while most participants who used the Baseline kept monitoring the required groups, i.e., \textit{``It is not that hard for me to monitor a group (with the same patterns) of students, but it would be impossible if there are multiple groups to track.''} (P4). As P3 suggested, with the notification system, instructors \textit{``could set alerts so that (they) don't need to look at those conditions all the time.''}

Nine participants mentioned the notification (with or without suggestion) would be helpful to \textit{``teach a 100/200 people programming class''}. \textit{``(\sys{}) Makes it much more convenient and feasible so you don't have to keep checking.''} (P10).

A lower level of temporal demand ($Median = 1.00, \sigma = 0.51$) was reported with the notification suggestions feature, compared to using the \sys{} without suggestion ($Median = 2.00, \sigma = 1.47$), however, this difference was not statistically significant.

\subsubsection{\sys{} Enables Instructors to be Prepared Before Sending Support}
Participants (P2, P3, P6, P9) also found that based on the meaning of the constraints applied to the alert, they could make more sense of how to help students with different types of difficulties. 
P6 suggested that \textit{``The topics filter can help professors/TAs prepare beforehand to help students instead of just going up to them and asking what they're struggling on.''}, while P9 also expressed \textit{``having different alerts to monitor different groups of students' behavior, so I can tell TA who can offer help in those different situation.''}

\section{Discussion}
Our study revealed several interesting insights about the automatic creation of notifications to assist with the monitoring of student progress in large classes. These insights related to instructor workflows, instructor decision making, and long-term student benefits of using such systems.

\subsection{\sys{}'s Impact on Instructor Workflows}
Instructors often tend to follow their mental model to set specific constraints to identify students' behavior in peer instruction sessions. However, instructors' strategies and workflows may vary from the monitoring system's intelligence level. For instance, in our baseline system, participants had to manually browse and process most of students' behaviors and interactions, while in the \sys{} without suggestion, participants began by using abstraction data (e.g., \textit{Conversation Topic}, \textit{Activity Level}). Based on the study results, when the system starts to generate things beyond summarization, e.g., providing users with suggestions, instructors' strategies to identify students can also be influenced.

This is consistent with findings in Explainable AI, which reported that the misalignment of decision-making criteria can affect users' levels of reliance on AI's recommendations~\cite{bansal2020does}. Ideally, we want instructors using \sys{} to perform better than either AI or themselves alone, yet this may be elusive due to the dynamic context in a classroom setting. Our study findings suggest that instructors are influenced directly and indirectly by exposure to the recommended notifications while still maintaining their own control over creating them. This provides new perspectives on human-AI collaboration in real-time data analytics tasks.

\subsection{Supporting Instructors' Decision Making}
Beyond tracking students' particular behavior and learning barriers, \sys{} might also provide instructors with a way to make sense of students' issues and suggest a solution regarding students' issues.
Prior work has shown that observable signals such as code completion or test case pass rates may not always align with students' actual understandings during PI~\cite{porter2011peer}. 
\sys{}'s recommended notification adapts to students' history data using students' behavior level information, providing instructors with insights from different perspectives.

\subsection{Long-Term Student Benefits}
\sys{} provided an opportunity for instructors to create diverse notifications and probe behavior patterns that were different from the instructors' common scope of behavior identification. This is helpful because prior work has suggested that with diverse mental models, students will often encounter new mistakes even on the same coding problem~\cite{meta_learning}. \sys{} shows that it can support instructors in identifying alternative learning patterns by adapting to students' behaviors in real time, helping students to be more likely to receive personalized feedback.

\subsection{Limitations}
There are several limitations to our user study. While we used authentic code and peer discussion data, we conducted the study in a simulated setting rather than in a real living classroom, which reduced the psychological intensity and potential for distraction among participants. Despite a high level of participants-reported usefulness and increases in both quantity and diversity of notifications using \sys{} in our lab study, instructors did not validate/interact with real students. Future research can also investigate instructors' processes for validating AI-generated content, particularly student-related information, in real-time classroom settings.

The discussion and submission data is only captured from one session during one class, which means that the result may not be generalizable to other contexts. Likewise, our evaluation of \sys{} used a dataset from a large programming lecture, limiting the scope to large-scale data. We suspect that participants spent less time on pattern identification and verifying notifications when dealing with smaller datasets, as there was less information to process. Future research can further explore the relationship between data size and visualization complexity.

Additionally, there are some limitations to our system. First, the choice of collaborative learning aspects used for generating notifications was limited to \textit{Pass Rate}, \textit{Activity Level}, \textit{Code Issue}, \textit{Conversation Topic}, and \textit{Team Structure}. Second, it is difficult to scale the number of notifications to help users understand and navigate the system. Moreover, the use of LLM could raise several concerns regarding stability, transparency, and trust. Though our approach adopted prior work's methods~\cite{Dai2023ChatGPTRS,10.1145/3636555.3636905predictingchallenge} that was examined through metrics (e.g., F1, NDCG@K), further evaluation of LLMs' capabilities in learning analytics is necessary. Lastly, our approach to generating notification suggestions faces limitations in handling context length when there are numerous messages or submissions. This issue could be mitigated with context management techniques in future work.

\section{Conclusion}

In this paper, we introduced \sys{}, a novel system that helps programming instructors create notifications, such as alerts and trackers, to better manage students' in-class Peer Instruction activity. 
It achieves this by leveraging LLMs to adapt time-sensitive contextual information, such as historical data changes, and recommends urgent notification units for instructors to monitor. Through our comparison study, we found that with our notification recommendations, participants discovered patterns that they were previously unaware of, and that the created notifications covered a more diverse range of collaborative learning metrics that led to changes in their creation strategies. 
Our work contributes new understandings and design lessons to real-time, large-scale collaborative learning analytics for domain experts who are novice users of Visual Analytics. Collectively, this opens a new door for building systems that support in-class collaborative learning at scale.

\bibliographystyle{ACM-Reference-Format}
\bibliography{ref}
\newpage

\appendix
\section{Appendix A: Prompts}
Below are the prompts used to generate the suggested notifications.
\subsection{Rank Groups by Issue Seriousness}
\textbf{System Prompt:} \\
\noindent \texttt{You are an Instructor tasked with 
identifying struggling groups with potential 
performance/communication issues and ranking them based on 
the seriousness of their issues from the given JSON data 
below about student groups' recent activities around solving 
the programming problem.} 

The output format is in JSON type:

\bigskip

\noindent \textbf{Programming Problem:} Write a function called \texttt{under100} that accepts a list of integers and returns the number of values in the list that are less than 100.

\bigskip

\noindent \textbf{Task:}
\begin{enumerate}
    \item Analyze each group's submission history, message history, and current status and identify the issues the group has. For instance, issues can be: "no active conversation", "insufficient submission attempts", "keep discussing a problem but no increase in pass rate", "members were asking for help but did not receive any from teammates", or "the main content of the conversation is not related to class", etc.
    \item Rank groups based on the seriousness of their issues based on the content of issues and groups' current performance (pass rate and activity level). For every group, you also need to record the aspect that you think the group has the most serious trouble with from the following list: ["pass rate", "amount of related messages in the conversation ", "topic of conversation", "member's participation in discussion"]. At last, summarize the issues you identified in [Task 1] in one short sentence. (Make sure all groups in the input are included in the output)
\end{enumerate}

\bigskip

\noindent \textbf{Input Format (JSON):}
A collection containing group objects (the key of the object is group's id). For each group, there are several components:

\begin{verbatim}
-currentStatus: [object] containing information about
    group's current status, including:
  -groupPassRate: [numeric] group's average passrate
  -teamActivity: [numeric] group's activity level
  -membersParticipatedNum: [numeric] number of members
    participated in the discussion
-topic: [string]: the summarized topic of the group discussion

-teamMembers: [Array] a list containing members' ID

-submissionHistory: [Array] a list of code submission records
from members of the group, each record containing:
  -time: [numeric] time the submission was made (second)
  -student_id: [string] the id of student who made the 
  submission
  -results: [boolean] whether the code submission has passed 
  the test
  -errorType: [string] type of the error of the code 
  submission
  -errorMessage: [string] error message of the code 
  submission
  -groupPassRate: [numeric] group's average passrate after 
  this submission

-messageHistory: [Array] a list of message records in the 
group, each containing:
  -time: [numeric] time the message was sent (second)
  -message: [string] the content of the message
  -sender_id: [string] the id of the student who sent the 
  message
  -senderActivityLevel: [numeric] the student's activity 
  level at that time
  -senderPassRate: [numeric] the pass rate of the sender 
  when the message was sent
  -activity: [string] the category of this message
  -topic: [string] the summarized topic of the group
    discussion when the message was sent
  -currentActivityLevel: [numeric] group's team activity 
  level at that time
  -currentPassRate: [numeric] group's pass rate at that time
\end{verbatim}

\bigskip

\noindent \textbf{Input Example (JSON):}
\begin{verbatim}
{"groupHistory":[
"qfNSCzEM1adKuYg8fS6s": {
                "currentStatus": {
                    "groupPassRate": 33.333333333333336,
                    "teamActivity": 0,
                    "membersParticipatedNum": 0,
                    "topic": "No Conversation"
                },
                "teamMembers": [
                    "kcufXPSXQdUrxgHMv5lh",
                    "RLF72ACbWtDW1b0DzQ15",
                    "Yvodndj3W2Ig8EdcymB6"
                ],
                "submissionHistory": [
                    {
                        "time": 68,
                        "student_id": "kcufXPSXQdUrxgHMv5lh",
                        "result": "not pass",
                        "errorType": "TypeError",
                        "errorMessage": "'int' object is not 
                        subscriptable",
                        "groupPassRate": 33.333333333333336
                    },
                    {
                        "time": 99,
                        "student_id": "kcufXPSXQdUrxgHMv5lh",
                        "result": "not pass",
                        "errorType": "SyntaxError",
                        "errorMessage": "invalid syntax",
                        "groupPassRate": 33.333333333333336
                    }
                ],
                "messageHistory": []
            },
            "rDJkBCxbE5NAdFSlWApx": {
                "currentStatus": {
                    "groupPassRate": 33.333333333333336,
                    "teamActivity": 3.6666666666666665,
                    "membersParticipatedNum": 2,
                    "topic": "Correcting Function 
                    Implementation for Counting"
                },
                "teamMembers": [
                    "rMInc3JASsCmwXFN6ZKH",
                    "VN8tUFi6ZCcXye9S2Nxw",
                    "oAKti7KYiXSKIanUBONs"
                ],
                "submissionHistory": [],
                "messageHistory": [
                    {
                        "time": 21,
                        "message": "where is a count used?",
                        "sender_id": "rMInc3JASsCmwXFN6ZKH",
                        "senderActivityLevel": 2,
                        "senderPassRate": 0,
                        "activity": "help-seeking",
                        "topic": "Correcting Function 
                        Implementation for Counting",
                        "currentActivityLevel": 
                        2.3333333333333335,
                        "currentPassRate": 33.333333333333336
                    },
                    {
                        "time": 27,
                        "message": "or what for I mean",
                        "sender_id": "rMInc3JASsCmwXFN6ZKH",
                        "senderActivityLevel": 3,
                        "senderPassRate": 0,
                        "activity": "help-seeking",
                        "topic": "Correcting Function 
                        Implementation for Counting",
                        "currentActivityLevel": 
                        2.6666666666666665,
                        "currentPassRate": 33.333333333333336
                    },
                    {
                        "time": 78,
                        "message": "it initializes your 
                        count value so the code knows where 
                        to start",
                        "sender_id": "oAKti7KYiXSKIanUBONs",
                        "senderActivityLevel": 5,
                        "senderPassRate": 100,
                        "activity": "help-giving",
                        "topic": "Correcting Function 
                        Implementation for Counting",
                        "currentActivityLevel": 3,
                        "currentPassRate": 33.333333333333336
                    },
                    {
                        "time": 115,
                        "message": "wait so is it asking for 
                        a list of the numbers or the amount 
                        of numbers that are below 100",
                        "sender_id": "rMInc3JASsCmwXFN6ZKH",
                        "senderActivityLevel": 4,
                        "senderPassRate": 0,
                        "activity": "help-seeking",
                        "topic": "Correcting Function 
                        Implementation for Counting",
                        "currentActivityLevel": 
                        3.3333333333333335,
                        "currentPassRate": 33.333333333333336
                    }
                ]
            },
]}

\end{verbatim}

\bigskip

\noindent \textbf{Output Format (valid JSON):}
A list of ranked group objects (Make sure to include all groups in the input), each containing:

\begin{verbatim}
- rank: The rank of this group based on the seriousness of 
its issues
- id: [string] The id of this group
- aspect: [string] The aspect that you think the group has 
the most serious trouble with
- issue: [string] A short-sentence summary of the group's 
issues identified in [Task 1].
\end{verbatim}

\bigskip

\noindent \textbf{Output Example (valid JSON):}
\begin{verbatim}
{"rankedGroupList":[
    {
        "rank": 1,
        "id": "qfNSCzEM1adKuYg8fS6s",
        "aspect": "amount of related messages in the 
        conversation",
        "issue": "No active conversation despite submission 
        attempts."
    },
    {
        "rank": 2,
        "id": "rDJkBCxbE5NAdFSlWApx",
        "aspect": "member's participation in discussion",
        "issue": "Members were asking for help but the 
        conversation lacks depth in problem-solving."
    }
]
}

\end{verbatim}
\subsection{Rank Students by Issue Seriousness}
\textbf{System Prompt:}
\noindent \texttt{You are an Instructor tasked with identifying struggling students with potential performance/communication issues and ranking them based on the seriousness of their issues from the given JSON data below about students' recent activities around solving the programming problem.  } 

The output format is in JSON type:

\bigskip

\noindent \textbf{Programming Problem:} Write a function called \texttt{under100} that accepts a list of integers and returns the number of values in the list that are less than 100.

\bigskip

\noindent \textbf{Task:}
\begin{enumerate}
    \item 1. Analyze each student's submission history, message history, and current status and identify the issues the group has. For instance, issues can be: "low pass rate", "insufficient discussion with group members", "insufficient submission attempts", "keep having the same code issue for student's code submissions ", "student was asking for help but no increase in pass rate", or "the main content of the student's messages is not related to class", etc.

    \item Rank students in descending order based on the seriousness (an issue with high seriousness may require you to send a TA to support, while an issue with low seriousness can simply be handled by the student who has that issue) of their issues based on the content of issues and students' current performance (pass rate and activity level). For every student, you also need to record the aspect that you think the student has the most serious trouble with from the following list: ["pass rate", "amount of related messages in the conversation ", "topic of conversation", "code issue"]. At last, summarize the issues you identified in [Task 1] in one short sentence. (Make sure all students in the input are included in the output)
\end{enumerate}

\bigskip

\noindent \textbf{Input Format (JSON):}
A collection containing student objects (the key of the object is student's id). For each student, there are several components:

\begin{verbatim}
-currentStatus: [object] containing information of student's 
current status, including:
  -passRate: [numeric] student's passrate
  -teamActivity: [numeric] student's activity level
  -topic: [string]: the summarized topic of the group 
  discussion

-submissionHistory: [Array] a list of code submission 
records of the student, each record containing:
  -time: [numeric] time the submission was made (second)
  -passRate: [numeric] pass rate of the student's code 
  submission
  -errorType: [string] type of the error of the code 
  submission
  -errorMessage: [string] error message of the code 
  submission

-messageHistory: [Array] a list of message records of the 
student, each containing:
  -time: [numeric] time the message was sent (second)
  -message: [string] the content of the message
  -activity: [string] the category of this message (6 
  categories in total: ["help-giving", "help-seeking", 
  "exchanging information and feedback", "mutual 
  encouragement and challenging", "joint reflection on 
  progress and process", "Not Class Related"])
  -currentTopic: [string] the summarized topic of the 
  student's group discussion when the message was sent
  -currentActivityLevel: [numeric] student's team activity 
  level at that time
  -currentPassRate: [numeric] student's pass rate at that 
  time
\end{verbatim}

\bigskip

\noindent \textbf{Input Example (JSON):}
\begin{verbatim}
{"studentHistory":{
            "0gL8b8z4viC8SXQiQi6x": {
                "currentStatus": {
            "passRate": 0,
            "teamActivity": 1.3,
            "topic": "Repeated Greetings and Minimal 
            Progress"
        },
        "submissionHistory": [],
        "messageHistory": [
            {
                "time": 71,
                "message": "wassup",
                "activity": "not related to the class",
                "currentTopic": "No Conversation",
                "currentActivityLevel": 0,
                "currentPassRate": 0
            },
            {
                "time": 90,
                "message": "same ",
                "activity": "help-seeking",
                "currentTopic": "Repeated Greetings and 
                Minimal Progress",
                "currentActivityLevel": 0.3,
                "currentPassRate": 0
            }
        ]
            },
            "DrMqavekheeqmxbSCSeg": {
                "currentStatus": {
            "passRate": 25,
            "teamActivity": 0.3,
            "topic": "Repeated Greetings and Minimal 
            Progress"
        },
        "submissionHistory": [
            {
                "time": 59,
                "passRate": 25,
                "errorType": "Logical Error",
                "errorMessage": ""
            },
            {
                "time": 82,
                "passRate": 25,
                "errorType": "Logical Error",
                "errorMessage": ""
            },
            {
                "time": 103,
                "passRate": 25,
                "errorType": "Logical Error",
                "errorMessage": ""
            },
            {
                "time": 113,
                "passRate": 25,
                "errorType": "Logical Error",
                "errorMessage": ""
            }
        ],
        "messageHistory": [
            {
                "time": 63,
                "message": "hi",
                "activity": "not related to the class",
                "currentTopic": "No Conversation",
                "currentActivityLevel": 0,
                "currentPassRate": 25
            }
        ]
            }
  }
}
\end{verbatim}

\bigskip

\noindent \textbf{Output Format (valid JSON):}
A list of descendingly ranked student objects (Make sure to include all students in the input), each containing:

\begin{verbatim}
- rank: The rank of this student based on the seriousness of 
his/her issues (an issue with high seriousness may require 
you to send a TA to support, while an issue with low 
seriousness can simply be handled by the student who has 
that issue)
-id: The id of this student (make sure the id appears in the 
input)
-aspect: The aspect that you think the student has the most 
serious trouble with from the following list: ["pass rate", 
"amount of related messages in the conversation ", "topic of 
conversation", "code issue"]
-issue: [string] A short-sentence summary of the group's 
issues identified in [Task 1].
\end{verbatim}

\bigskip

\noindent \textbf{Output Example (valid JSON):}
\begin{verbatim}
{"rankedStudentList":[
    {
        "rank": 1,
        "id": "DrMqavekheeqmxbSCSeg",
        "aspect": "code issue",
        "issue": "Repeated logical errors in code 
        submissions and lack of progress despite consistent 
        pass rate."
    },
    {
        "rank": 2,
        "id": "0gL8b8z4viC8SXQiQi6x",
        "aspect": "passrate",
        "issue": "No conversation with group members and 
        minimal progress despite seeking help."
    }
]
}
\end{verbatim}
\subsection{Extracting Group Code Issues:}
\textbf{System Prompt:}
\noindent \texttt{You are a programming 
Instructor tasked with summarizing student 
groups' common performance/ communication issues 
from the given JSON data below about student 
groups' issues while solving the programming 
problem. }

\bigskip

\noindent \textbf{Programming Problem:} Write a function called under100 that accepts a list of integers and returns the number of values in the list that are less than 100.

\bigskip

\noindent \textbf{Task:}
\begin{enumerate}
    \item Summarize the common issues based on issues and aspects from the input in one short sentence.

    \item In the summarized issue description you have in [Task 1], identify which of the following aspects was described.
\end{enumerate}

\bigskip

\noindent \textbf{Input Format (JSON):}
A list of issue objects containing:

\begin{verbatim}
-issue: [string] content of 
performance/communication issues groups have 
while working on the programming problem.
-aspect: [string] The aspect that the group has 
the most serious trouble with. ( List of 
aspects: ["pass rate", "amount of related 
messages in the conversation ", "topic of 
conversation", "member's participation in 
discussion"])

\end{verbatim}

\bigskip

\noindent \textbf{Input Example (JSON):}
\begin{verbatim}
{"groupIssueList":[
{
                "aspect": "pass rate",
                "issue": "Lowest pass rate with 
                multiple error types and no 
                conversation."
            },
            {
                "aspect": "pass rate",
                "issue": "Very low pass rate and 
                no active conversation."
            },
            {
                "aspect": "pass rate",
                "issue": "Low pass rate with 
                minimal conversation and 
                participation."
            },
            {
                "aspect": "amount of related 
                messages in the conversation",
                "issue": "No active conversation 
                despite some submission 
                attempts."
            },
            {
                "aspect": "amount of related 
                messages in the conversation",
                "issue": "No conversation 
                despite submission attempts."
            },
  ]
}
\end{verbatim}

\bigskip

\noindent \textbf{Output Format (valid JSON):}

\begin{verbatim}
-issueSummary: [string] The short-sentence 
summary of common issues in the input issue 
reflecting the common aspect of issues.
-aspectList: [array] each containing identified 
aspects in the issueSummary. Candidate aspects: 
["pass rate", "amount of related messages in the 
conversation ", "topic of conversation", 
"member's participation in discussion"]
\end{verbatim}

\bigskip

\noindent \textbf{Output Example (valid JSON):}
\begin{verbatim}
{“summary”:{
  "issueSummary": "Low pass rates paired with 
  inadequate or no conversation",
  "aspectList": ["pass rate", "amount of related 
  messages in the conversation"]
}
}
\end{verbatim}
\subsection{Extracting Individual Code Issues:}
\textbf{System Prompt:}
\noindent \texttt{You are a programming 
Instructor tasked with summarizing student 
groups' common performance/ communication issues 
from the given JSON data below about student 
groups' issues while solving the programming 
problem. }

\bigskip

\noindent \textbf{Programming Problem:} Write a function called under100 that accepts a list of integers and returns the number of values in the list that are less than 100.

\bigskip

\noindent \textbf{Task:}
\begin{enumerate}
    \item Summarize the common issues based on issues and aspects from the input in one short sentence.

    \item In the summarized issue description you have in [Task 1], identify which of the following aspects was described.
\end{enumerate}

\bigskip

\noindent \textbf{Input Format (JSON):}
A list of issue objects containing:

\begin{verbatim}
A list of issue objects containing:
-issue: [string] content of 
performance/communication issues 
students have while working on the 
programming problem.
-aspect: [string] The aspect that the 
student has the most serious trouble 
with. ( List of aspects: ["pass rate", 
"amount of related messages in the 
conversation ", "topic of 
conversation", "code issue"])
\end{verbatim}

\bigskip

\noindent \textbf{Input Example (JSON):}
\begin{verbatim}
{"studentIssueList":[
{
                "aspect": "code issue",
                "issue": "Repeated 
                TypeError issues and 
                seeking help without 
                improvement."
            },
            {
                "aspect": "amount of 
                related messages in 
                the conversation",
                "issue": "Repeated 
                Logical Errors and not 
                related messages 
                despite seeking help."
            },
            {
                "aspect": "pass rate",
                "issue": "Struggling 
                with final test cases 
                and actively seeking 
                help."
            },
            {
                "aspect": "pass rate",
                
                "issue": "Improved 
                pass rate through 
                submissions but still 
                seeking help."
            },
            {
                "aspect": "code issue",
                "issue": "Consistent 
                Logical Errors in 
                submissions."
            },
  ]
}

\end{verbatim}

\bigskip

\noindent \textbf{Output Format (valid JSON):}

\begin{verbatim}
-issueSummary: [string] The short-
sentence summary of common issues in 
the input issue reflecting the common 
aspect of issues.
-aspectList: [array] each containing 
identified aspects in the 
issueSummary. Candidate aspects:["pass 
rate", "amount of related messages in 
the conversation ", "topic of 
conversation", "code issue"]
\end{verbatim}

\bigskip

\noindent \textbf{Output Example (valid JSON):}
\begin{verbatim}
{
  "summary": {
    "issueSummary": "Issues with code 
    errors and pass rate while 
    struggling with efficient 
    communication.",
    "aspectList": ["code issue", "pass 
    rate", "amount of related messages 
    in the conversation"]
  }
}
\end{verbatim}

\section{Appendix B: Formative Study Tasks}
\label{formative_study_task}
\begin{itemize}
    \item Task 1: Identify groups with specific patterns regarding progress in solving the exercise and participating in group discussion,
    \item Task 2: Identify students who were seeking help and had unsolved issues,
    \item Task 3: Track and understand multiple groups' dynamics in a real-time setting,
    \item Task 4: Understand groups' common issues.
\end{itemize}

\end{document}